\documentclass[%
reprint,
superscriptaddress,
nofootinbib,
amsmath,amssymb,
aps,
prl,
]{revtex4-1}
\usepackage{xr-hyper}
\usepackage{hyperref}
\usepackage{graphicx} 
\usepackage{dcolumn}
\usepackage{bm}
\usepackage{braket}
\usepackage{dsfont}
\usepackage{color,colortbl}
\usepackage[table,xcdraw]{xcolor}
\usepackage{multirow}
\usepackage{subcaption}
\usepackage{mwe}
\usepackage{booktabs}
\usepackage{caption}
\usepackage{lipsum}
\usepackage{babel,blindtext}
\usepackage{amsmath}
\usepackage[toc,page]{appendix}
\usepackage[symbol*]{footmisc}
\usepackage{float}

\usepackage{natbib}
\usepackage{filecontents}

\makeatletter

\newcommand*{\addFileDependency}[1]{
\typeout{(#1)}
\@addtofilelist{#1}
\IfFileExists{#1}{}{\typeout{No file #1.}}
}\makeatother

\newcommand*{\myexternaldocument}[1]{%
\externaldocument{#1}%
\addFileDependency{#1.tex}%
\addFileDependency{#1.aux}%
}

\myexternaldocument{SI}

\begin{document}

\title{Unraveling the Catalytic Effect of Hydrogen Adsorption \\on Pt Nanoparticle Shape-Change}

\author{Cameron J. Owen$^{*,\dagger}$}
\affiliation{Department of Chemistry and Chemical Biology, Harvard University, Cambridge, Massachusetts 02138, United States}

\author{Nicholas Marcella$^{*,\dagger}$}
\affiliation{Department of Chemistry, University of Illinois, Urbana, Illinois 61801, United States}

\author{\\Yu Xie}
\affiliation{John A. Paulson School of Engineering and Applied Sciences, Harvard University, Cambridge, Massachusetts 02138, United States}

\author{Jonathan Vandermause$^{\ddagger}$}
\affiliation{John A. Paulson School of Engineering and Applied Sciences, Harvard University, Cambridge, Massachusetts 02138, United States}

\author{Anatoly I. Frenkel}
\affiliation{Department of Materials Science and Chemical Engineering, Stony Brook University, Stony Brook, New York 11794, United States}
\affiliation{Chemistry Division, Brookhaven National Laboratory, Upton, New York 11973, United States}

\author{Ralph G. Nuzzo}
\affiliation{Department of Chemistry, University of Illinois, Urbana, Illinois 61801, United States}

\author{Boris Kozinsky$^{\dagger,}$}
\affiliation{John A. Paulson School of Engineering and Applied Sciences, Harvard University, Cambridge, Massachusetts 02138, United States}
\affiliation{Robert Bosch LLC Research and Technology Center, Watertown, Massachusetts 02472, United States}

\def\thefootnote{$*$}\footnotetext{These authors contributed equally.}\def\thefootnote{\arabic{footnote}}

\def\thefootnote{$\ddagger$}\footnotetext{Currently at D.E. Shaw Research.}\def\thefootnote{\arabic{footnote}}

\def\thefootnote{$\dagger$}\footnotetext{Corresponding authors\\C.J.O., E-mail: \url{cowen@g.harvard.edu}\\N.M., E-mail: \url{nmarcella@bnl.gov}\\B.K., E-mail: \url{bkoz@seas.harvard.edu}\def\thefootnote{\arabic{footnote}}}

\newcommand\bvec{\mathbf}
\newcommand{\mathsc}[1]{{\normalfont\textsc{#1}}}

\begin{abstract}
The activity of metal catalysts depends sensitively on dynamic structural changes that occur during operating conditions.
The mechanistic understanding underlying such transformations in small Pt nanoparticles (NPs) of $\sim1-5$ nm in diameter, commonly used in hydrogenation reactions, is currently far from complete.
In this study, we investigate the structural evolution of Pt NPs in the presence of hydrogen using reactive molecular dynamics (MD) simulations and X-ray spectroscopy measurements.
To gain atomistic insights into adsorbate-induced structural transformation phenomena, we employ a combination of MD based on first-principles machine-learned force fields with extended X-ray absorption fine structure (EXAFS) measurements. 
Simulations and experiments provide complementary information, mutual validation, and interpretation.
We obtain atomic-level mechanistic insights into `order-disorder' structural transformations exhibited by highly dispersed heterogeneous Pt catalysts upon exposure to hydrogen.
We report the emergence of previously unknown candidate structures in the small Pt NP limit, where exposure to hydrogen leads to the appearance of a `quasi-icosahedral' intermediate symmetry, followed by the formation of `rosettes' on the NP surface.
Hydrogen adsorption is found to catalyze these shape transitions by lowering their temperatures and increasing the apparent rates, revealing the dual catalytic and dynamic nature of interaction between nanoparticle and adsorbate.
Our study also offers a new pathway for deciphering the reversible evolution of catalyst structure resulting from the chemisorption of reactive species, enabling the determination of active sites and improved interpretation of experimental results with atomic resolution.
\end{abstract}

\maketitle

\section{Introduction}
In heterogeneous catalysis research, deciphering reaction mechanisms typically involves a combination of experimental characterization with chemical intuition to narrow down the myriad of potential reaction pathways and active site geometries that could facilitate the surface-molecule interactions that govern the chemical process \cite{Marcella2022}. 
While this approach is tried-and-true for a set of relatively simple scenarios involving small molecules on ideal static surfaces, the advent of intricate subnanometer heterogeneous catalysts, coupled with a growing understanding of their dynamic behavior \cite{Li2021}, has cast doubt on the efficacy of conventional approaches for discerning active sites within such catalysts. 
Specifically, the complexity incurred via dynamic restructuring of the catalyst adds to the problem, leading to a quick breakdown of applicable chemical intuition. 
In this work, we use an intuition-free and data-driven approach to disentangle the mechanisms of reaction-induced restructuring of Pt nanocatalysts.
First, \textit{ab initio} trained reactive machine-learned force fields (MLFFs) are used to evolve nanoparticle systems, resulting in quantum-mechanically accurate simulations over experimentally-relevant time- and length-scales. 
Thus, realistic active site geometries are revealed in an unbiased fashion rather than constrained by intuition. 
Subsequently, extended X-ray absorption fine structure (EXAFS) spectra are calculated on these structures and compared to experimentally observed EXAFS spectra, resulting in the simultaneous validation of the underlying theory and explanation of the experimental EXAFS in terms of local structure parameters from which the underlying atomistic mechanisms can be dissected. 
This intuition-free approach will be especially useful in the modeling of heterogeneous catalysts.

Heterogeneous nanoparticles (NPs) are difficult to characterize due to the configurational complexity introduced by interparticle heterogeneity (\emph{e.g.}, the particle size and shape distribution) and intraparticle heterogeneity (\emph{e.g.}, the coexistence of truncated and variously coordinated structures of surface terraces, edges, and vertices). 
This complexity is magnified by the tendency of NPs to restructure in response to chemisorption of adsorbate coupled with changes in temperature or pressure. 
In Pt NPs, metal-adsorbate interactions during pretreatment and operating conditions have been shown to cause significant restructuring, as evidenced by CO and H$_2$ exposures \cite{Li2021,small_influence_2012}. 
For instance, CO adsorption can lead to a flattening of supported Pt NP structures, causing distortion of the bond length distribution and general loss of crystalline order, while hydrogen adsorption reduces surface strain due to charge transfer effects \cite{small_influence_2012}. 
L. Li et al. reported an `order-disorder' transformation in Pt NPs exposed to H$_2$ at high temperatures, which was found to be dependent on particle size \cite{li_noncrystalline--crystalline_2013}. 
Using time-resolved TEM imaging, Y. Li et al. observed dynamic structural changes, \emph{i.e.} a loss in the atomic order of the Pt lattice sites, in ceria-supported NPs over minutes that were directly related to the conditions of the water gas shift reaction, particularly high partial pressures of CO \cite{Li2021}. 
F. D. Vila et al. discovered low-frequency dynamics (nuclear motion with frequencies exceeding 1 THz and dynamic structural disorder arising from slower frequencies in the $0-1$ THz range) in Pt NPs that led to a significant macro-scale effect over timescales of $10$ ps, specifically a negative thermal expansion coefficient resulting from the temperature-dependent interaction of the NP with the support \cite{vila2017anomalous}. 

\begin{figure}
\includegraphics[width=\columnwidth]{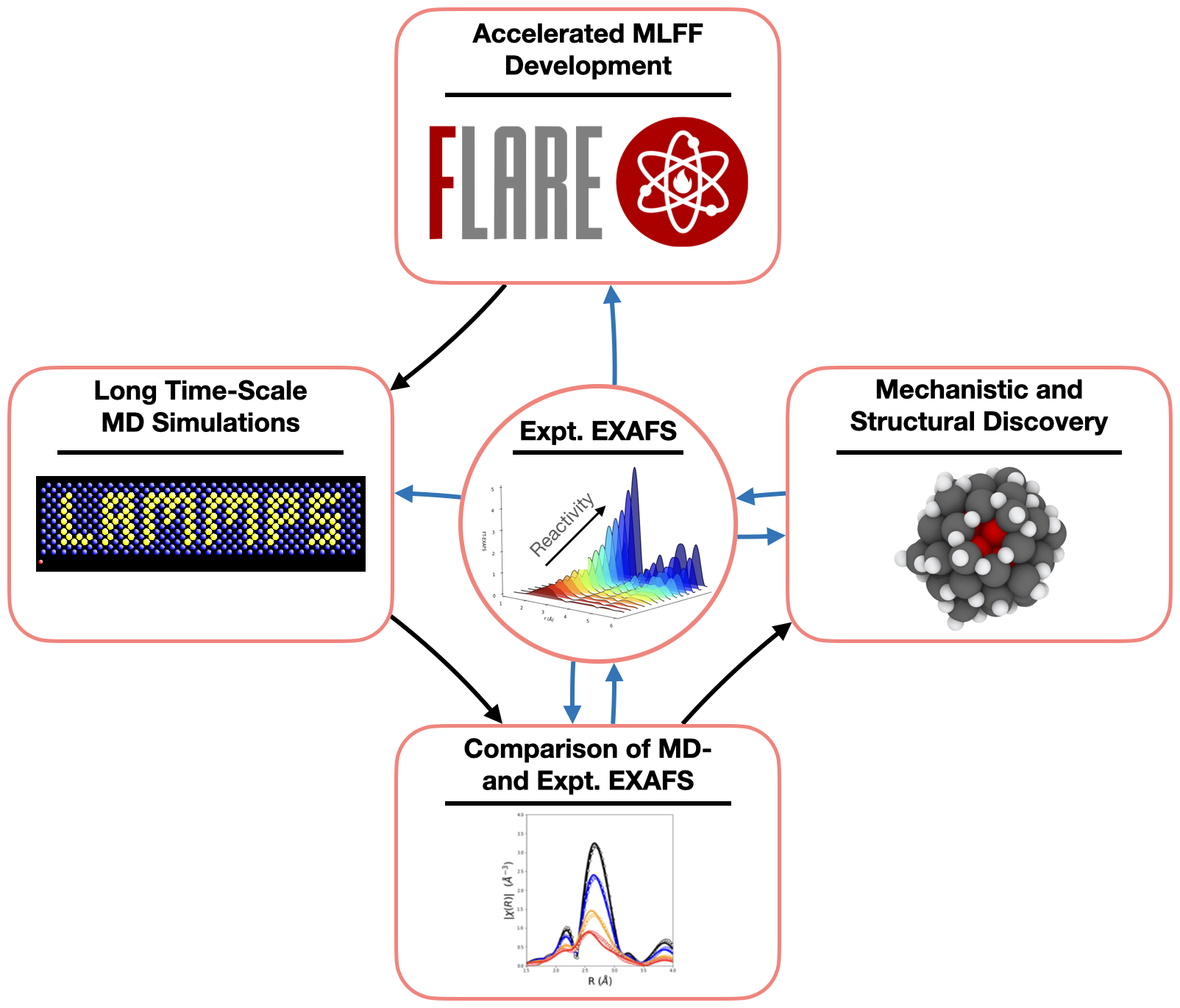}
\caption{Schematic of the combined simulation and spectroscopic framework for studying nanoparticle structural transformations under inert and reactive conditions. 
(\textbf{Center}) Experimental EXAFS is used to inform multiple components of the workflow.
(\textbf{Top}) Machine-learned force fields (MLFFs) are constructed with active learning from \textit{ab initio} data using the FLARE code.
(\textbf{Left}) MLFFs are validated according to Pt material properties and Pt-H interactions and used in long time-scale MD simulations. 
(\textbf{Bottom}) MD trajectories are generated using the MLFF and used to calculate the MD-EXAFS. 
The MD-EXAFS is then compared to experimental EXAFS of well-defined systems collected over various temperatures and reactive conditions. 
(\textbf{Right}) The validated model can be used for extraction of atom-level mechanistic insights in systems that are not well understood experimentally.}
\label{fig:toc}
\end{figure}

In traditional and contemporary approaches to understanding the structure-performance relationship of heterogeneous catalysts, experimental data acquired through spectroscopy \cite{C0CS00089B,doi:10.1021/nl500553a,doi:10.1021/acs.jpcc.2c05929,https://doi.org/10.1002/cctc.201500688}, scattering \cite{Sun2017,doi:10.1021/acs.jpcc.1c10824,doi:10.1021/jacs.2c13666}, imaging \cite{doi:10.1021/acscatal.8b01321,Yan2016,C4FD00035H}, or a combination of these techniques \cite{doi:10.1021/acscatal.2c03863,GREAVES1985203,Ijima2002,PhysRevB.85.195419} are employed to refine the range of potential active site geometries for consideration in static \textit{ab initio} calculations. 
EXAFS spectroscopy, arguably the most critical and widely used method for examining the structural evolution of functional nanomaterials \cite{newville2014fundamentals, Ijima2002, osti_1787395}, is commonly employed to obtain short-range `local' structural parameters that are used to develop a rough geometrical model. 
Such a model can be used as a starting point for density functional theory (DFT) calculations, resulting in a significantly reduced configurational space.

Nevertheless, the information acquired through EXAFS is severely limited by the fact that it is a time- and configurational-average over all absorbing atoms and therefore all particles of various sizes and shapes in the ensemble. 
Even for a hypothetical system containing identical particles, the recovery of atom-specific information from EXAFS would be prohibited by Nyquist criterion \cite{PhysRevB.48.9825}. 
Consequently, despite consistent advancements in spatial \cite{osti_1787395} and temporal resolution \cite{Muller:rv5040} and data analysis techniques over recent years, EXAFS is inherently limited in its capacity to directly probe atomistic dynamics in individual NPs. 
One intriguing possibility to bypass the information limit of this spectral inversion problem is to formulate a forward modeling approach. 
Here, we utilize molecular dynamics (MD) simulations, for which the EXAFS spectrum is computed for atomic structures generated by an MD trajectory \cite{kas2021advanced}. 
If the force field used for MD is accurate, and the experimentally observed EXAFS signal agrees with that obtained by the MD simulation, then the atomistic details captured by the simulation can be used to extract the full structural and mechanistic information. 

This approach, referred to as MD-EXAFS \cite{doi:10.1021/jp960160q,doi:10.1063/1.471711}, has been used to directly validate MD simulations obtained with empirical force fields (FFs) \cite{ANSPOKS20112604,PhysRevB.85.075439,Cicco_2002,PhysRevB.83.115409} and \textit{ab initio} calculations \cite{C3SC50614B,doi:10.1021/acsnano.5b00090,doi:10.1146/annurev-anchem-061318-114929,BOCHAROV2020109198}. 
The case of reactive systems has remained challenging to address in this way due to the short time- and length-scale limitations of \textit{ab initio} techniques and the absence of accurate reactive empirical FFs. 
Recent developments of surrogate MLFFs have allowed for atomic forces, total energies, and stresses to be learned as functions of atomic positions, meaning that the resulting model can accurately approximate \textit{ab initio} calculations with much higher computational efficiency. 
These MLFFs are then used to perform MD simulations at near-\textit{ab initio} accuracy. 
For instance, ML MD was used to generate structures for EXAFS analysis of pure bulk metals \cite{SHAPEEV2022111028}. 
Until recently, however, MLFFs for heterogeneous reactive systems have not been demonstrated, and consequently, such analysis of NP catalysts using EXAFS data has not been reported but could provide a lens with which to interpret the atomistic evolution of these complicated systems.
Over the last several years, MLFFs trained on \textit{ab initio} calculations have exhibited substantial improvement in accuracy and efficiency relative to previous empirical FF methods \cite{doi:10.1021/acs.jpclett.1c01204,https://doi.org/10.1002/adma.201902765}. 
First demonstrations have been able to provide insight into atomistic mechanisms of heterogeneous reactions at experimentally relevant length- and time-scales, wherein a fully dynamic model can describe molecule-surface interactions and reactions without any assumptions \cite{Vandermause2022,Lim2020EvolutionDynamics}. 

Unlike empirical FFs using constrained functional forms, MLFFs are built on non-parametric architectures and trained on large \textit{ab initio} datasets, meaning that they are flexible and can describe more diverse atomic environments, including bond-breaking at \textit{ab initio} accuracy. 
The current state-of-the-art MLFF models are also increasingly robust in transferability tasks, \emph{i.e.}, generalize well outside their training sets, as has been demonstrated for both kernel and deep equivariant neural network methods \cite{Kovacs2021LinearRMSE,Batzner2021E3-EquivariantPotentials,Musaelian2023,owen2023complexity}. 
Importantly, when considering catalysts under reaction conditions with possible structural transformations, it is difficult to construct diverse and relevant training sets which ensure the accuracy of the ML model across such broad composition and configurational spaces. 
To address this, a subset of MLFFs has been purposefully developed to quantify uncertainties of their predictions on desired physical observables (\emph{i.e.}, energies, forces, and stresses), enabling highly data-efficient and automated training of the models via active learning \cite{Vandermause2020,Vandermause2022,https://doi.org/10.48550/arxiv.2211.09866}. 
The utility of such Bayesian MLFFs and active learning was recently demonstrated for constructing a model for the hydrogen evolution reaction over Pt(111) \cite{Vandermause2022}, which was able to capture the kinetics of the catalytic process in close agreement with experimental measurements. 
That MLFF also achieved record speed and size for reactive ML MD simulations, scaling up to 0.5 trillion atoms \cite{Johansson2022Micron-scaleLearning}, but only explored the idealized Pt(111) surface, omitting the aforementioned complications associated with diverse atomic environments on NPs and their responses to environmental stimuli. 

In this work, we examine the dynamic structure of Pt NPs under inert and reactive hydrogen atmospheres at \textit{ab initio} accuracy using MLFFs at time-scales greatly exceeding those accessible with \textit{ab initio} methods. 
We accomplish this by building on recently introduced Bayesian MLFFs for describing H/Pt heterogeneous interactions in combination with X-ray spectroscopy for structural characterization. 
Crucially, our MLFF is validated with both theoretical and experimental data, giving credence to subsequent predictions made by the MLFF during MD simulations. 
First, we use the underlying \textit{ab initio} method, DFT at the PBE level, across several benchmarks, \emph{e.g.}, Pt bulk elastic constants, equation of state, and the adsorption of H and H$_2$ on various Pt facets. 
Second, the MLFF is also validated via comparison of its predictions of structural dynamics to available experimental EXAFS data.
This comparison provides an excellent target for interpretation of the atomistic dynamics of these systems, as EXAFS can be collected across a broad range of temperatures, pressures, and reaction conditions, serving as an easily accessible and powerful tool for combination with MLFFs.
This scheme allows for direct atomistic interpretation of NP systems of realistic complexity under reaction conditions without the requirement of expert chemical intuition or assumptions in the modeling task.
Ultimately, we demonstrate that the introduction of hydrogen can either stabilize or destabilize close-packed Pt structures depending on NP size, and we explain this observation via the underlying quantum-mechanical interactions in the systems, which the MLFF is able to capture.
Such a demonstration of the combined use of ML MD supported by spectroscopy pushes the field closer to unraveling the atomistic mechanisms governing catalytic systems with realistic complexity, importantly without bias, on adequate time- and length-scales.

\begin{figure*}
\centering
\includegraphics[width=\textwidth]{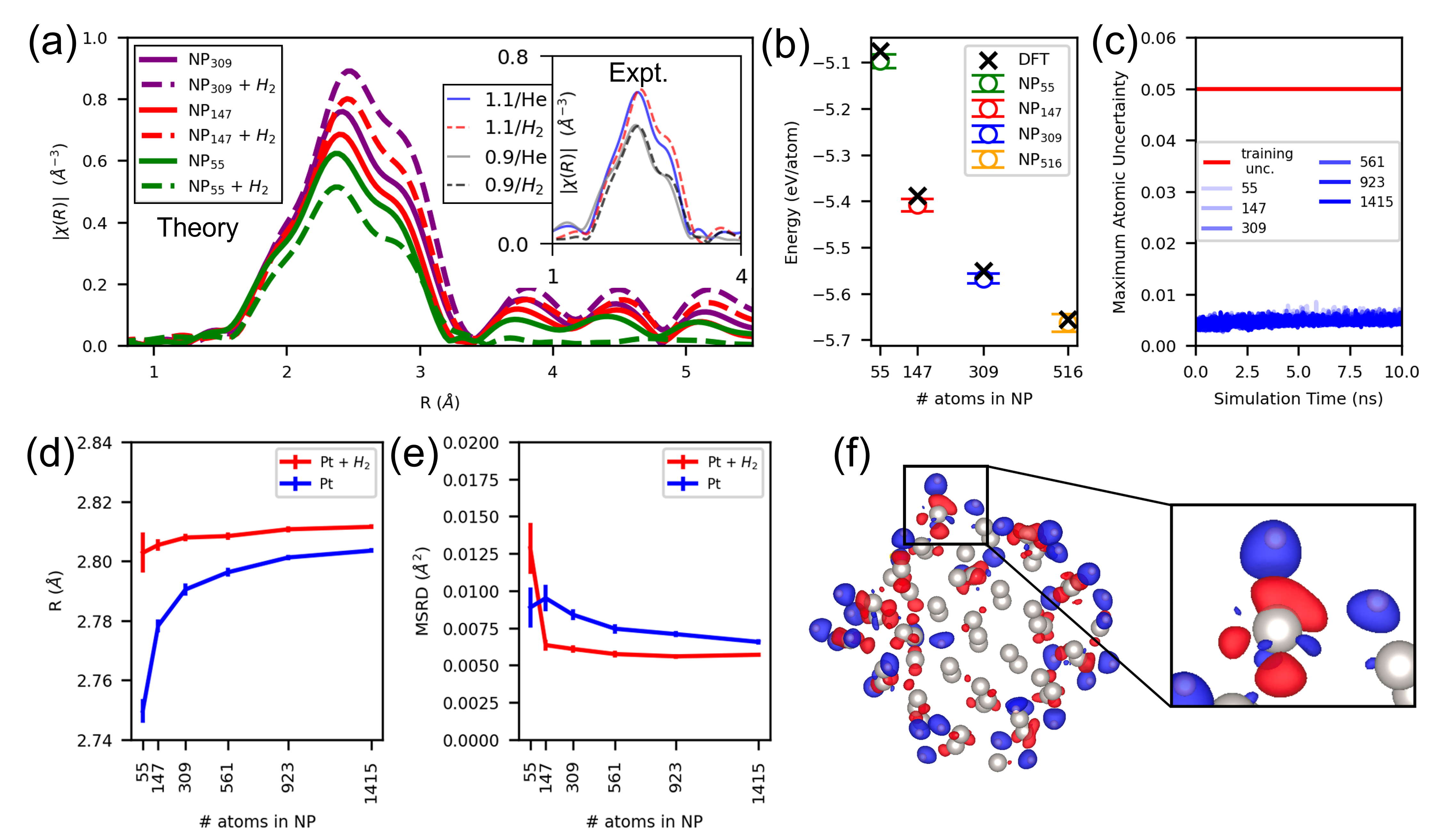}
\caption{(\textbf{a}) Simulated FT-MD-EXAFS for cuboctahedral Pt NPs ranging in size from $55$ to $309$ atoms both with and without $\sim$ 0.1 bar H$_2$. 
We note that MD-EXAFS calculations for NPs with more than $147$ atoms were completed with a surrogate neural network (described in the Methods section and SI). 
The inset shows four experimental FT-EXAFS spectra corresponding to the labels 1.1/He, 1.1/H$_2$, 0.9/H$_2$, and 0.9/H$_2$ in which the particle size in nanometers is followed by the gas environment ($\sim$ 0.1 bar).
(\textbf{b}) Comparison of MLFF predicted cohesive energies per atom of Pt cuboctahedral NPs as a function of size. 
Open circles denote MLFF predictions, whereas black crosses are DFT. 
The $99\%$ confidence interval of each prediction by the MLFF is provided. 
Only the $55$ and $147$ atom NPs are included in the training set, but despite this, the MLFF demonstrates excellent transferability to the $309$ and $561$ atom particles. 
(\textbf{c}) Maximum atomic uncertainty of ML MD predictions made by the MLFF as a function of MD simulation time under exposure to H$_2$ at 0.1 bar. 
(\textbf{d}) Average interatomic distance R and (\textbf{e}) MSRD for nanoparticles from $55$ to $1415$ atoms with (red) and without (blue) a H$_2$ atmosphere. 
(\textbf{f}) Charge density difference isosurface from a $55$ atom simulation frame under exposure to H$_2$. Pt atoms are silver, and H atoms are pink. 
Blue isosurfaces denote a gain in charge density, while red denote a loss.}
\label{fig:deploy}
\end{figure*} 

\section{Results}
\subsection{Quantifying NP Shape and Disorder with ML MD and EXAFS}
This section presents an investigation of the restructuring of Pt NPs, as evidenced by the distribution of Pt-Pt bonds. 
The complementary approaches, ML MD simulations and EXAFS are used to analyze the Pt-Pt radial distribution function (RDF), focusing on the first nearest neighbor distances within the $2.0-3.4$ \AA{} range. 
The integrated area, mean, and variance of the RDF correspond to the EXAFS-derived structural parameters: coordination number (CN), average interatomic distance (R), and mean squared relative displacement of the interatomic distance (MSRD) (also referred to as the Debye-Waller factor or disorder in conventional EXAFS fitting), respectively. 
The MSRD is especially useful in our case, as the term combines the static (configurational) and dynamic (vibrational) disorders of the X-ray absorbing atoms and their nearest neighbors. 
All of these quantities are sensitive to the surface-area-to-volume (SAV) ratio of NPs due to the differences between surface, subsurface, and bulk bonds. 
Trivially, bonds on or near the surface contribute more to the overall signal when fewer bulk bonds are included in the ensemble average. 
ML MD simulations and MD-EXAFS data are thus utilized to investigate the restructuring of nanoparticles (NPs) with and without the presence of H$_2$. 
The capability of ML MD simulations to capture global trends manifested by the SAV ratio of NPs \cite{PhysRevB.81.115451, frenkel_view_2001} and the distribution of Pt-Pt interatomic distances \cite{small_influence_2012}, is assessed. 
These effects are examined using ML MD across various particle sizes and shapes, including `magic-number' particles with cuboctahedral, icosahedral, and ino-truncated decahedral symmetries containing 55 to 1415 atoms. 
The results are then compared with experimental observations from EXAFS and previous findings in the literature.

Before presenting the results, a few considerations must be made clear regarding comparison of the ML MD results to experimental NP systems. 
First, realistic NP catalysts exhibit a size distribution whose width can vary and must be considered for accurate interpretation \cite{rodriguez2013situ, doi:10.1021/acs.jpcc.3c00571}. 
Second, experimental methods for characterizing NP shape and faceting are qualitative; thus, NP samples are commonly assumed to exhibit a shape distribution with respect to the lowest-energy particle shapes, which are themselves size-dependent \cite{phan_what_2019}. 
Lastly, NPs are usually attached to or embedded in a support material, the effects of which are poorly understood and difficult to characterize but are postulated to influence particle size and shape distributions during both synthesis, pretreatment, and catalyst operation \cite{SanchezNonBulk}. 
Summarily, the geometries of NPs are not well understood experimentally.
The task of modeling these systems thus becomes drastically more difficult once additional degrees of freedom are included, namely through the inclusion of temperature and adsorbates.
Therefore, we cannot establish a direct correspondence between experimental EXAFS data and MD-EXAFS data, as there is not enough information regarding the exact sizes (\emph{i.e.}, the number of atoms within each particle) as well as the NP geometries. 
Instead, we leverage the sample-average relative signals of EXAFS for inert and H$_2$ conditions over Pt NPs to compare the MLFF predictions with experimental reference. 
Additionally, we consider only free-standing NPs in ML MD simulations, neglecting NP-support interactions. 
Explicit treatment of these interactions will be the focus of our follow up investigation. 

\subsection{Structural Effects of Hydrogen Adsorption from ML MD Agree with Experiment}
The experimental systems we consider are $0.9 \pm 0.2$ and $1.1 \pm 0.3$ nm Pt NPs, supported by $\gamma$-Al$_2$O$_3$ as studied previously in \cite{SanchezNonBulk}. 
Experimental EXAFS in that work was collected for these particles independently under He and $10\%$ H$_2$ at ambient pressure and temperature, following previous analysis \cite{SanchezNonBulk,Timoshenko2017b,Frenkel2012,RevModPhys.72.621}. 
The experimental Fourier transformed EXAFS (FT-EXAFS) for these particles is provided partially in the inset of Fig. \ref{fig:deploy}(a). 
The plotted data shows the influence of particle size and atmosphere on the spectra. 
In particular, the mean position and amplitude of the first peak, which are correlated with R and MSRD, respectively, change between particle size and atmosphere. 
The peak amplitude, associated with the coordination number, particle size, and disorder, varies between sizes due to the inclusion of undercoordinated surface atoms in the sample ensemble average, \emph{i.e.} an SAV effect. 
The difference in peak amplitude and position between atmospheres is thus isolated as an effect of hydrogen adsorption. 
It is important to note that the NP ensemble has a size distribution close to the `order-disorder' domain proposed in \cite{li_noncrystalline--crystalline_2013} and may contain highly disordered subnanometer particles. 
This is reflected in the experimental EXAFS through the MSRD values and visual confirmation with TEM \cite{li_noncrystalline--crystalline_2013}. 
To interpret the measured data, we employ the MLFF in MD to simulate NP systems within this size domain, examining the effects of size and H$_2$ exposure on EXAFS signatures and the atomic mechanisms governing this behavior.

\begin{figure*}
\centering
\includegraphics[width=\textwidth]{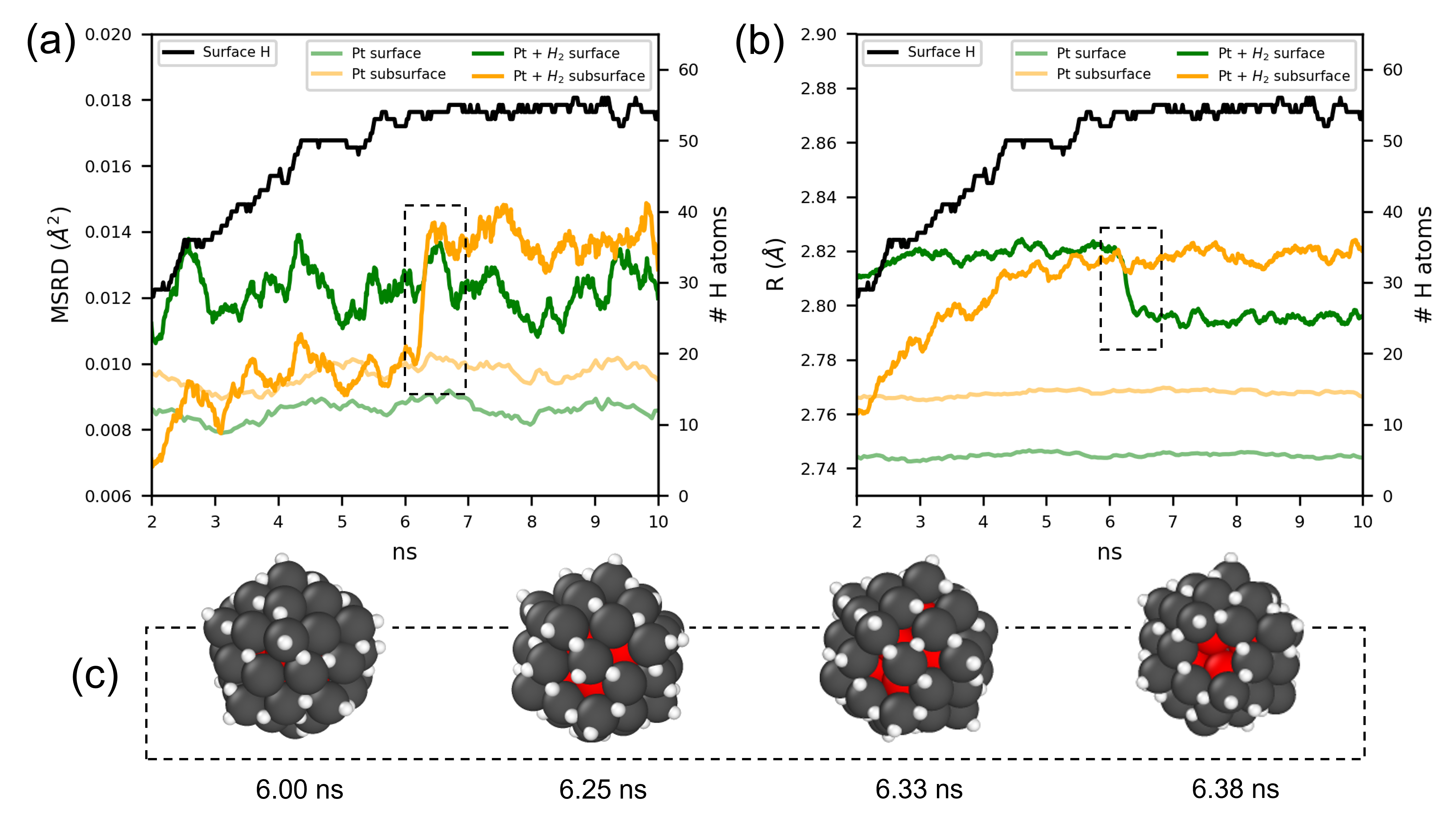}
\caption{Trajectory analysis of the $55$ atom cuboctahedral Pt particle with and without H$_2$. 
(\textbf{a}) Average MSRD as a function of simulation time (ns) with (bold traces) and without (light traces) H$_2$.
The traces are separated into Pt surface (green) and subsurface (orange) contributions to the total signal.
Simultaneously, H adsorption (black) is also plotted as a function of simulation time and the number of adsorbed H atoms is plotted on the right y-axis. 
(\textbf{b}) Average partial MSRD separated into surface (green) and subsurface (orange) components for the $55$ atom cuboctahedral particle with (bold) and without (light) H$_2$. 
(\textbf{c}) Structural snapshots from the morphological transition to yield a rosette, observed in the $55$ atom particle between 6 - 6.38 ns under high H$_2$ coverage. 
Surface Pt atoms are colored dark grey, subsurface Pt atoms are red, and adsorbed hydrogen atoms are white. 
The final structure at 10 ns yields multiple rosettes on opposite sides of the particle, as is shown in Fig. S15.
}
\label{fig:55cub}
\end{figure*} 


The Atomic Simulation Environment (ASE) \cite{Larsen_2017} was used to construct magic-number particles of $55$, $147$, and $309$ atoms with cuboctahedral symmetry to match the range of experimental size distributions between $0.9 \pm 0.2$ and $1.1 \pm 0.3$ nm.
Larger NPs containing 561, 923, and 1415 atoms were also considered to obtain general trends with respect to particle size. 
For simulations under hydrogen, H$_2$ molecules were introduced into the simulation cell until the pressure of $\sim 0.1$ bar was reached. 
The systems were then simulated according to the procedure outlined in the Methods section. 
The FT-EXAFS calculated from the ML MD trajectories (FT-MD-EXAFS) obtained from these trajectories are shown in Fig. \ref{fig:deploy}(a). 
For all NP sizes, the average interatomic distance (R) scales with the SAV ratio (surface bonds are shorter than bulk bonds, which are more heavily weighted in smaller particles) Fig. \ref{fig:deploy}(d), consistent with the literature \cite{PhysRevB.81.115451, frenkel_view_2001}.
In particle sizes larger than $55$ atoms, hydrogen exposure leads to an increase of the average Pt-Pt interatomic distance R and a reduction in the MSRD, while the coordination number does not change.
As shown in Fig. \ref{fig:deploy}(a), these experimental results are consistent with the MD-EXAFS prediction, providing a level of validation of the MLFF. 

Unlike the larger particles, exposure of the $55$ atom NP to H$_2$ increases both R and MSRD relative to the bare particle, and the coordination number changes significantly, pointing to significant overall restructuring. 
This result is unexpected, as hydrogen adsorption is typically observed to relax the surface atoms, decreasing the MSRD and increasing the Pt-Pt interatomic distances, as several experiments have reported \cite{small_influence_2012,frenkel2014critical,frenkel2013situ,SanchezNonBulk}.
The FT-MD-EXAFS reflects the aforementioned trends in the peak amplitudes and positions.
Because all particles maintain a constant number of atoms and coordination over the simulation time under both inert and H$_2$ atmospheres (Fig. S14), the changes in FT-MD-EXAFS amplitude can be interpreted in terms of the disorder (quantified by the MSRD in Fig. \ref{fig:deploy}(e)).
For example, in the $147$ and $309$ atom particles in Fig. \ref{fig:deploy}(a), hydrogen is seen to have an ordering effect on the structure, reflected by a decrease in the MSRD and subsequent increase in FT-MD-EXAFS amplitude. 
This observation is consistent with experimental EXAFS measurements of $\sim 1$ nm NPs and greater, where the particles are large enough that chemisorbed H$_2$ tends to stabilize the initial structure \cite{Surnev_2001,doi:10.1021/cs3004006}. 

So far, the emergence of expected trends in NPs larger than 55 atoms is encouraging and serves as additional preliminary validation of the MLFF. 
To understand the ability of the MLFF model to simulate systems outside of the training set (\emph{e.g.}, larger NP sizes and symmetries other than cuboctahedral), the MLFF was also explicitly tested by evaluating bare cuboctahedral, icosahedral, and ino-truncated decahedral NP cohesive energies (Fig. \ref{fig:deploy}(b) and S16(a,e)) and maximum atomic local energy uncertainties observed during MD simulations (Fig. \ref{fig:deploy}(c) and S16(b,f)).
Excellent agreement is observed between the MLFF and DFT predictions across the NP size domain, and at the same time, no uncertainties exceeded the training threshold. 
These are both important observations, as only the $55$ and $147$ atom cuboctahedral NPs were included in the training set, but the MLFF can still accurately determine total energies, forces, and stresses of $309$ and $561$ atom systems, and this holds across NP shapes.

\subsection{Atomistic Insights into NP Responses to H$_2$}
We now provide an explanation for the hydrogen-induced elongation of Pt-Pt interatomic distances observed across all NPs considered as well as direct insight into the disparate structural response to this elongation based on particle size. 
Previous works have established charge-transfer mechanisms using DFT by which hydrogen chemisorption on extended Pt surfaces results in a transfer of charge from the \textit{d}-orbitals of Pt to the \textit{s}-orbitals of H \cite{catal8100450}. 
This leads to a weakening in the Pt-Pt interactions and outward relaxation of the surface layer Pt atoms, thus increasing the interatomic distances to their neighbors. 
This observation is in line with the delocalized bonding picture of metals, where an increase in the CN of a given atom typically results in a nonlinear increase in the length of the bonds with nearest neighbors \cite{doi:10.1098/rsta.1991.0021}. 
To directly verify this charge-transfer hypothesis, several ML MD frames at high coverage of H$_2$ on Pt$_{55}$ were extracted, and static DFT calculations were performed (excluding all gas-phase hydrogen to reduce the simulation box dimensions for efficiency in DFT). 
These results are displayed in Fig. \ref{fig:deploy}(f), where the charge transfer between surface Pt and adsorbed H is apparent from the difference in isosurface values, where the charge is effectively redistributed from Pt to H. 
We see that the MLFF, through its high-dimensional description of \textit{ab initio} interatomic interactions in a local neighborhood, learns the effect that this charge transfer between the surface Pt atoms and the chemisorbed hydrogen has on the surface structure. 
This results in the correct prediction of shift in peak positions of the FT-MD-EXAFS of these NPs under H$_2$.
We posit that particle shape-change in the small NP limit is due to this charge transfer between Pt and the chemisorbed H, which influences the Pt-Pt bonds in the core of the NP as there is not enough bulk to effectively shield these interactions. 
Ultimately, this hypothesis is rather intuitive, as the larger the bulk reservoir, the smaller the distortions in interatomic distance, leading to reduced mobility of the atoms.
Hence, small Pt particles then have a higher propensity to change-shape under H chemisorption due to the overall weakening of Pt-Pt bonds through the actual core of the NP, whereas larger particles effectively gain more `bulk-like' bonding environments, due to a layer of chemisorbed hydrogen, the charge transfer effect sof which are quickly screened by the larger reservoir of bulk bonds.

To further understand the increased disorder observed in the $55$ atom particle under H$_2$, we also performed mechanistic analysis of the ML-MD simulation. 
We simulated the $55$ atom particle under $\sim$0.1 bar of H$_2$ at 256 K for a total duration of 10 ns in order to closely match experimental conditions, as explained in the Methods section.
The NP was initialized with cuboctahedral symmetry, and H$_2$ was only in the gas-phase (\emph{i.e.}, no chemisorbed H).
Full simulation details are also provided in the Methods section.
The average CN of atoms in the $55$ atom particle did not stay constant during this simulation under H$_2$ exposure, as opposed to the constant CN of larger particles at the same temperature and H$_2$ exposure.
As shown in Fig. S13 and S14, the average CN starts at $7.86$, corresponding to a $55$ atom particle with cuboctahedral symmetry \cite{doi:https://doi.org/10.1002/9781118844243.ch3}, then rapidly increases to $8.51$, corresponding to an icosahedron of $55$ atoms \cite{doi:https://doi.org/10.1002/9781118844243.ch3}, before returning to an average CN closer to that of a cuboctahedron, despite the particle symmetry not being cuboctahedral. 
The MSRD, average R, and breakdown of both surface and subsurface contributions to the bond topology are provided in Fig. \ref{fig:55cub}(a) and Fig. \ref{fig:55cub}(b), respectively.

Thus, starting from cuboctahedral symmetry at 256 K under the presence of 0.1 bar H$_{2}$, the particle quickly transitions into a \textit{quasi-icosahedral} geometry within the first 320 picoseconds under a moderate coverage of hydrogen ($\sim$ 12 chemisorbed atoms). 
The particle symmetry is characterized as quasi-icosahedral based on Polyhedral Template Matching \cite{Larsen_2016} analysis as implemented in the OVITO code \cite{ovito} denoting all atomic environments as disordered, but the particle exhibits ($111$) surface packing with five-fold symmetry. 
Moreover, the icosahedral symmetry for the bare particles is energetically lower in both 0 K relaxations using DFT and the MLFF and at finite temperature in ML MD for all of the magic numbers considered. 
This is corroborated by other theoretical investigations \cite{PhysRevLett.93.065502}. 
To the best of our knowledge, however, small icosahedral Pt nanoparticles have yet to be confirmed experimentally, which is challenging due to limitations in time- and length-scale resolutions that would be required for such a determination. 
After reaching maximum hydrogen coverage around 6 ns, as shown in Fig. \ref{fig:55cub}(c), a second structural transition is observed, evidenced by a sudden change in the subsurface MSRD and the Pt-Pt interatomic distances in the surface layer, as shown in Fig. \ref{fig:55cub}(a) and \ref{fig:55cub}(b). 

\begin{figure*}
\centering
\includegraphics[width=1.0\textwidth]{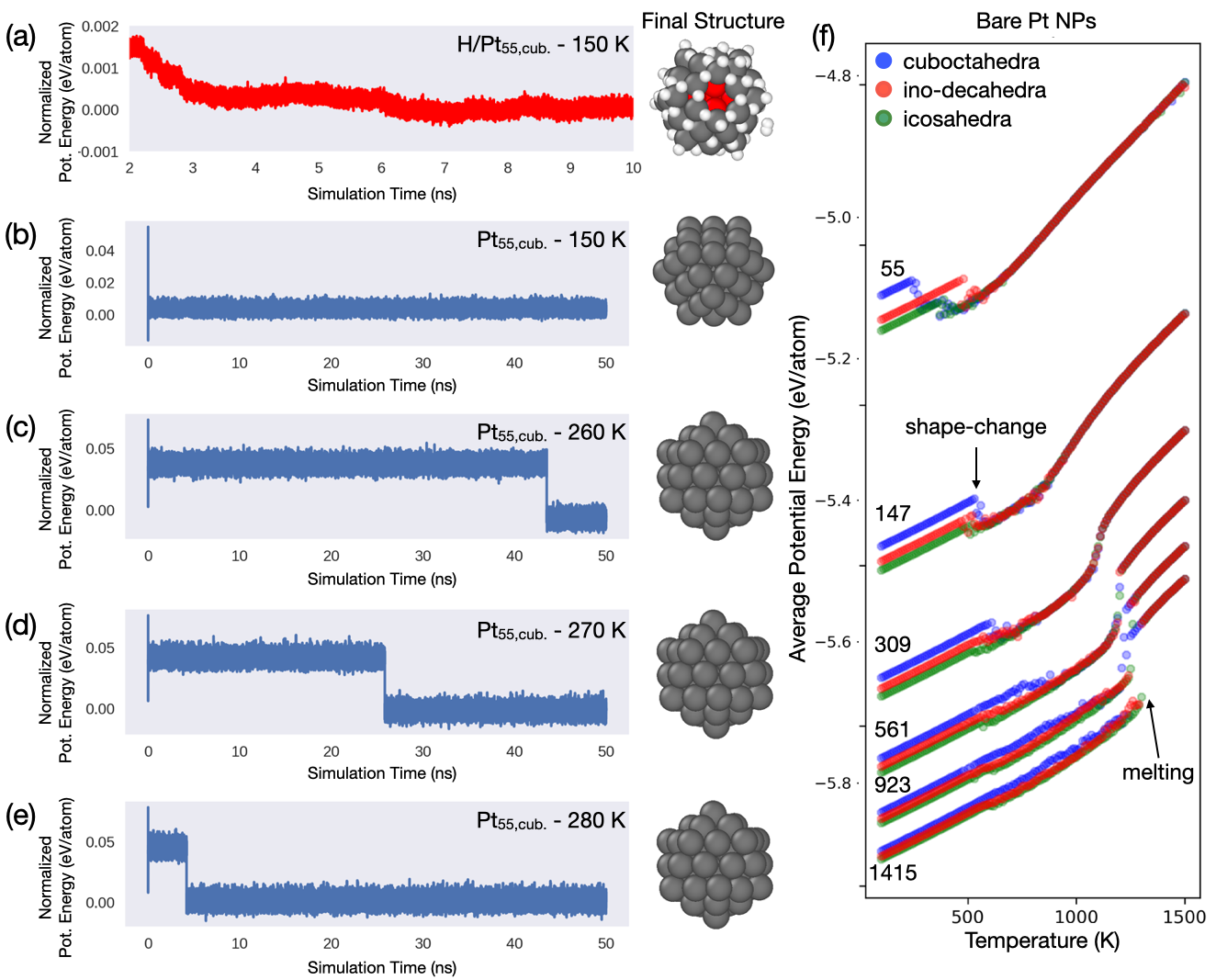}
\caption{
(\textbf{a}) Normalized potential energy (eV/atom) to the final potential energy of the simulation as a function of simulation time (ns) for a 55 atom cuboctahedral Pt NP  (Pt$_{55,cub.}$) under exposure to H$_{2}$ at 150 K. 
The final shape contains rosettes, as shown in the corresponding snapshot taken at the end of the simulation (10 ns).
(\textbf{b-e}) Normalized potential energy (eV/atom) for Pt$_{55,cub.}$ bare NPs as a function of simulation time at separate temperatures (150, 260, 270, and 280 K, respectively), where the shape-change can be observed at lower simulation times as temperature is increased.
Final NP snapshots are provided for each simulation, depicting cuboctahedral (\textbf{b}) or icosahedral symmetries (\textbf{c-e}).
(\textbf{f}) Average potential energy (eV/atom) as a function of simulation temperature for bare Pt NPs. 
Each point represents an independent trajectory, where the potential energy is averaged over the final 5 ns of the total simulation time (total time of each simulation is 50 ns).
Domains for both NP shape-change and melting as a function of particle shape and size can be observed, as well as the relative energies between particle symmetries.}
\label{fig:transfer}
\end{figure*} 

Ultimately, the 55 atom NP adopts a quasi-ordered geometry where hexagonal rings form on the surface next to five-fold coordinated sites, shown in Fig. \ref{fig:55cub}(c). 
As the simulation progresses further in time, more rings form at these five-fold sites on other sides of the particle, effectively creating single-atom vacancies in the surface layer of the particle, or \textit{rosettes} as coined by Apr\`a et al. \cite{PhysRevLett.93.065502}, on the particle surface, decorated by rings of chemisorbed hydrogen. 
The formation of hexagonal rings at five-fold sites has previously been studied for magic number Pt and Au NPs under the assumption of a shape-change mechanism unique to the icosahedral symmetry \cite{PhysRevLett.93.065502}. 
In this mechanism, rosettes form at five-fold sites on the NP surface, and DFT at the PBE level (the same as employed here) predicts these structures to be unstable for bare Pt, Ag, and Pd NPs but stable for Au. 
For Pt, however, local optimizations of the 55 atom structure after the introduction of multiple rosettes lead to closure of the rosette, which yields low-symmetry and low-energy particles that are in competition with the high-symmetry icosahedron in terms of stability.
In the H-Pt case reported here using our MLFF, we see that the multiple-rosette geometry is the most stable structure, thus; while hydrogen does increase the disorder of the 55 atom particle relative to its initial symmetry, it also increases the stability of this novel structure.

Thus, our study presents NPs containing rosettes as new geometries that can occur in small Pt NPs, with the requirement that a sufficient amount of chemisorbed hydrogen is present to stabilize this structure.
This result was realized under 0.1 bar of hydrogen and in less than 10 ns of simulation time, which suggests that these structures would likely be present at even lower pressures of H$_2$ at experimental time-scales.
To reiterate, the shape transformation begins from cuboctahedral symmetry, and the formation of rosettes only occurs after a quasi-icosahedral particle emerges due to a small coverage of chemisorbed hydrogen. 
In conclusion, we propose that the quasi-icosahedral and rosette geometries account for the observed persistence of disordered sub-nanometer Pt NPs reported by L. Li et al. \cite{li_noncrystalline--crystalline_2013}. 

\subsection{Catalysis of NP Shape via H$_2$ Exposure is Dependent on Initial Particle Symmetry}
Given the response of the 55 atom cuboctahedral NP to both temperature and hydrogen exposure and the identification of nontrivial structures under H$_2$, we then examined how these size-dependent behaviors translate across particle shapes. 
We investigated the temperature dependence of the shape-changes of the 55 atom and larger cuboctahedral, icosahedral, and ino-decahedral particles in both bare and H-covered configurations. 
First, we wanted to understand the effect of temperature on the shape-change of the bare 55 atom cuboctahedral Pt NP to change-shape to yield the icosahedron, since this system is very well characterized from the previous sections.
Potential energy predictions and the corresponding final shapes of each temperature and adsorbate exposure for Pt$_{55}$ in its cuboctahedral symmetry are shown in panels (a - e) of Fig. \ref{fig:transfer}. 
We observe a drastic difference in the behavior of the NP shape as a function of temperature and H$_2$ exposure, where H adsorption reduces the thermal requirement to change shape by over 100 K compared to the bare uncovered configuration.
This is corroborated by panels Fig. \ref{fig:transfer}(a) and (b), in comparison with panel (f), which shows the average potential energy predictions across a wide range of temperatures, shapes, and sizes.
Comparing the temperature requirement for the bare, cuboctahedral NP with 55 atoms of $\sim$250 K to change shape, as indicated by the cusp in potential energy in panel (f), panel (a) clearly demonstrates a catalytic effect of H$_2$ exposure on the shape-change to yield rosettes at 150 K, and also much shorter time-scales.  
As for time-scale of each shape-change, panels (b - e) of Fig. \ref{fig:transfer} demonstrate a change in shape by the reduction in potential energy as a function of simulation time, which systematically shifts to shorter times as temperature is increased.
This points to shape-change in small Pt NPs being kinetically limited, which means that both environmental and support effects could play large roles in facilitating or prohibiting such changes.

As for the other particle shapes, subtle differences in the behavior of icosahedral and ino-truncated decahedral particles are observed in Fig. \ref{fig:transfer}(f).
Comparison of the bare particle average potential energies provides insight into the stable NP shapes as a function of temperature across particle size, as well as the resulting symmetry after shape-change.
We see that the icosahedron is the most stable out of the magic number particles considered, and that shape-change to this symmetry occurs at markedly higher temperatures as particle size increases.
However, we also observe that the icosahedral symmetry exhibits a cusp, indicating potential shape change at increased temperatures.
Upon inspection, the cusp in the icosahedral series corresponds to an amorphous transition, \emph{e.g.} surface pre-melting, which preceeds particle melting.
Only once the particles are large enough ($\geq$ 923 atoms), the cusp corresponding to shape-change becomes less apparent, due to the presence of a large bulk reservoir, which constrains the particle against changing shape until particle melting occurs.

\begin{figure}
\includegraphics[width=1.0\columnwidth]{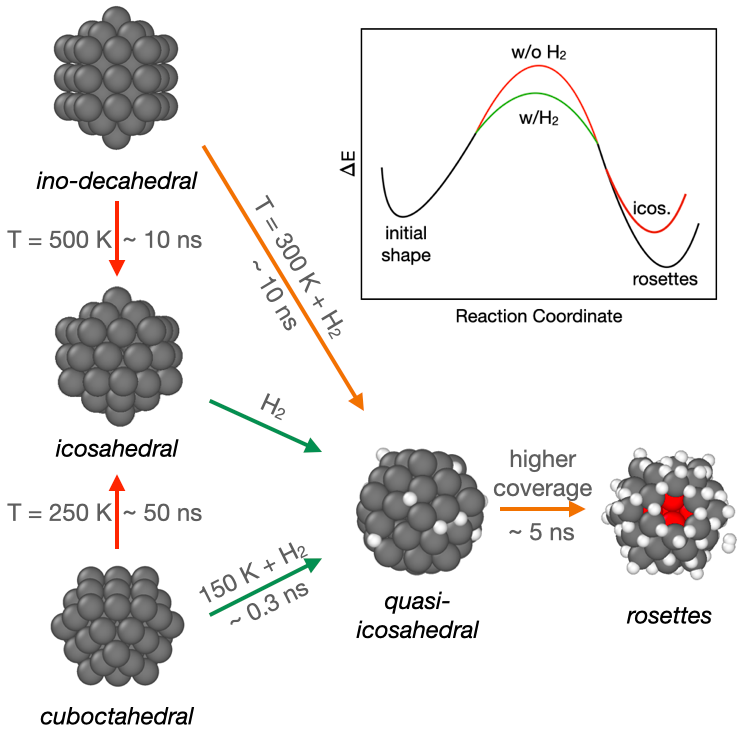}
\caption{
Overview of the structural transition pathways with respect to input shape (\textbf{left}), with corresponding conditions (for the 55 atom particles) between them. Each input shape transitions to yield icosahedral symmetry if the conditions allow. The transition through quasi-icosahedral and eventual appearance of rosettes is only made possible through H$_2$ exposure. The catalytic effect of H$_2$ on the 55 atom cuboctahedron is also made apparent by the transition temperatures and times (\textbf{bottom left}). (\textbf{Upper right}) Schematic illustrating the catalytic effect of H$_2$ exposure on NP shape-change.
}
\label{fig:last}
\end{figure}

As for particle exposure to H$_2$, the induced ordering-effect of H$_2$ is seemingly absent in the icosahedral NPs, whereas a similar effect to the cuboctahedra is observed in the decahedral NPs, as shown in Fig. S16.
Similar to the $55$ atom cuboctahedral particle, the $55$ atom icosahedral particle exhibits a much larger MSRD under H$_2$ relative to inert conditions. 
This is because H$_2$ adsorption pushes the cuboctahedral particles first into the quasi-icosahedral and subsequently the rosette geometry.
Thus, we make an important observation here, in that the icosahedral and cuboctahedral particles act similarly since they progress through the quasi-icosahedral state, as was shown in the left-most snapshot of Fig. \ref{fig:55cub}(c), but the initial barrier is reduced by construction for the icosahedral simulations, so it can more quickly reach the rosette-containing structure.

So, there is a seemingly important intermediate structure in the shape-change mechanism in the small-NP limit, identified here as the quasi-icosahedral state. 
This hypothesis is supported by the fact that the $55$ atom decahedral particle does not exhibit the same increase in MSRD and retains its initial morphology throughout the H$_2$ simulation at low temperatures.
However, if the simulation temperature is increased to either $\sim 540$ or $\sim 300$ K for the bare or H$_2$ exposed 55 atom ino-decahedral particles, they also undergo structural transitions to yield quasi-icosahedral or rosette geometries, respectively.
The dependence on initial particle symmetry on transitions towards these intermediate and final geometries is portrayed in Fig. \ref{fig:last}, where the simulation conditions and time-scales are provided from each input structure.
To further illustrate the effects of hydrogen on catalyzing these structural changes, a characteristic energy-reaction coordinate diagram is included.
As particle size increases, however, the influence of hydrogen exposure on shape-change to yield the icosahedral symmetry and rosettes reduces dramatically.
This was directly observed for the 147 atom cuboctahedral particle, which did not change shape under exposure to 0.1 bar of hydrogen at temperatures lower than 540 K, which was the shape-change temperature determined for the bare particle to yield the icosahedral symmetry.
We explain this via reduced bonding strength of H at higher temperatures, meaning that if the bare particle transition temperature is around the desorption temperature of H$_2$ ($\sim$ 500 K \cite{MILLER1993395}) this catalytic effect will be greatly diminished, if not eliminated entirely.

Lastly, we verify that these transitions are reversible by taking both the quasi-icosahedral and rosette structures, removing hydrogen, and simulating each at 256 K, as was done for direct comparison to the experimental EXAFS data in \cite{SanchezNonBulk}. 
We find that the `quasi-icoshedral' NP relaxes to yield a particle with icosahedral symmetry, and the rosette closes to resemble the quasi-icoshedral intermediate structure.
It is also observed that this quasi-icoshedral structure closely matches the one proposed in Fig. 3 of \cite{PhysRevLett.93.065502} by Apr\`a et al.
Hence, reversibility is established between the rosette containing structure and particles with the quasi-icosahedral intermediate shape, illustrating the influence of hydrogen on `catalyzing' the shape of small Pt NPs.

\section{Discussion}\label{sec:disc}
This work provides a previously inaccessible atomistic understanding of the dynamic structure of Pt NPs, specifically their dynamic structural response to reactive hydrogen atmospheres.
The comprehensive morphological study of this phenomenon underscores the efficacy of coupling MLFF accelerated MD simulations with MD-EXAFS and experimental EXAFS to understand and predict atomistic mechanisms occuring at tens of nanoseonds across a variety of nanoparticles ranging in size from 55 to 1415 atoms, all at \textit{ab initio} accuracy.
This source of validation for the relative effects of temperature and chemisorption of reactants allows for creedence in the subsequent atomistic mechanisms underlying these phenomena, as predicted by the MLFF.
The changes in particle morphology observed in the simulations closely mimic the trends of experimental EXAFS data, where Pt NPs respond dynamically to H$_2$ exposure, varying as a function of particle size.  

We leverage the computational efficiency and \textit{ab initio} accuracy of the MLFF to predict potential morphologies of small Pt NPs, thus assisting the interpretation of experimental EXAFS data. 
Specifically, we postulate the importance of initial particle geometry in providing access to the intermediate quasi-icosahedral and final rosette-containing geometries, illustrated by the differences in the evolution of the initially cuboctahedral, icosahedral, and ino-truncated decahedral particles as a function of temperature and exposure to H$_2$. 
The outcome of the MLFF-driven MD simulation of Pt NPs in hydrogen aligns well with the size dependence of the `order-disorder' transition in the small NP-limit observed experimentally.

Importantly, direct simulations, validated by experiment, yield insights into the nanosecond timescales associated with chemically-catalyzed structural transformations -- information impossible to obtain experimentally. 
We find that shape-change of the 55 atom particles is effectively `catalyzed' via H$_2$ exposure, posited as a consequence of the absence of a bulk reservoir that can effectively screen the charge-transfer effects of H chemisorption, leading to a reduction in the temperature and time-scales at which the quasi-icosahedral intermediate appears, which then progresses to yield rosettes at high coverage of chemisorbed H$_2$. 
This new understanding has implications not only for the fundamental understanding of the size-dependent dynamic structure of ubiquitously important catalytic materials but also for devising novel strategies to design and manipulate their catalytic properties using reactive atmospheres at reduced temperatures.

The unprecedented combination of accuracy and scalability of the computational methodologies with experimental structural characterization employed here also opens avenues for examining dynamic behavior of a broad range of nanoparticle compositions, shapes, and sizes under a variety of environmental conditions such as temperature, pressure, and adsorbates. 
The resulting real-time direct atomistic information enables the quantitative analysis to decouple experimental EXAFS signals, augmenting previously challenging and tentative interpretations. 
Using atomistic information from EXAFS with ML MD also sets the stage for further investigations with explicit inclusion of NP supports and multiple adsorbates, moving the field closer to full-scale realistic models and hence, full \textit{ab initio} atomistic understanding of catalytic systems. 

\section*{Methods} \label{sec:methods}
\subsection{Machine-Learning Bayesian Force Field}
The machine-learned force field used for the HPt system is from the Fast Learning of Atomistic Rare Events (FLARE) open-source code: \url{https://github.com/mir-group/flare}. 
Briefly, the geometry of local atomic environments (within a cutoff radius $r_c$) is encoded in descriptors using the atomic cluster expansion (ACE) \cite{Drautz2019AtomicPotentials}.
The normalized dot-product kernel is used to measure the similarity between atomic descriptors, which is then used in the construction of a sparse Gaussian process regression model (SGP).
The SGP provides an inherent mechanism to quantify predictive uncertainties for atomic forces, energies, and stresses, which can be used by the active learning algorithm described in the next section to select \textit{ab initio} training data `on-the-fly' during an MD simulation. 
After the training is finished, the SGP force field is mapped to a polynomial model \cite{Vandermause2022,Xie2021BayesianStanene} for high-efficiency in production MD simulations, being much faster than the SGP and without losing accuracy. 

In the data collection stage with parallel active learning trajectories, as described in the next section, as well as the final offline training to get the master MLFF from the entire data set, the ACE-B2 descriptor was employed.
By using the second power of the normalized dot product kernel, `effective' 5-body interactions within each descriptor can be obtained, which is sufficiently complex for describing Pt with high accuracy \cite{owen2023complexity}.
To explore the effect of static model parameters, 5 frames of a H/Pt(111) active learning trajectory were used as a training set, and the remaining frames as the test set to perform a coarse grid search. 
Maintaining consistency in notation with the original work of ACE \cite{Drautz2019AtomicPotentials}, 
ultimately, we chose $n_{\text{max}}=8$ for radial basis, $l_{\text{max}}=3$ for angular basis, and the cutoff radius $r_{cut} = 4.25 \rm{\AA}$ for Pt-Pt, $3.0 \rm{\AA}$ for Pt-H, and $3.0 \rm{\AA}$ for H-H, which were found to yield the largest log-marginal likelihood while reducing the energy, force, and stress errors to their respective minima. 
These results are also consistent with those determined previously for H-Pt \cite{Vandermause2022,Johansson2022Micron-scaleLearning}.
Parity and the associated errors between the final MLFF and DFT on energy, force, and stress predictions are provided in the SI, along with several of the validation protocols considered.

\subsection*{Active Learning for Data Collection} \label{sec:flare}
The Bayesian active learning workflow is also from FLARE and has been described in detail elsewhere \cite{Vandermause2020,Vandermause2022}.
Briefly, the MLFF runs MD simulations with predictive uncertainties. 
When the uncertainty is high, a DFT calculation is made on the current MD frame.
Once high-uncertainty atomic environments are selected from the DFT frame and added to the sparse-set, the SGP is then re-mapped onto a lower-dimensional surrogate model, and the system is allowed to continue in the MD simulation.
Positions are then updated with respect to forces using the LAMMPS Nos\'e-Hoover NVT ensemble until another atomic environment is deemed as high uncertainty by the SGP \cite{Xie2023Uncertainty-awareSiC}. 
This workflow enables efficient exploration of the configuration space and facilitates rapid data collection of only relevant frames.
Hence, DFT is only called when the model deems an atomic environment as uncertain, making this method much more efficient than \textit{ab initio} MD, where a DFT calculation is required at every time-step (\emph{e.g.} $\Delta t = 5$ fs for pure Pt or $\Delta t = 0.1$ fs for H/Pt). 
Initial data for the HPt system were taken from our previous efforts in \cite{Vandermause2022} and \cite{Johansson2022Micron-scaleLearning}, which focused on the H$_2$ interaction with Pt(111), so the additional active learning described here focused on NPs.
The systems considered during parallel active learning simulations were: bulk Pt with and without tensile and compressive strains, the Pt(111) surface with and without H$_2$, gaseous H$_2$, and NPs of various sizes with and without H$_2$.
A summary of these systems and their active learning results is provided in the SI.

The threshold for calling DFT based on atomic uncertainties is defined at the onset of each active learning trajectory, and was set to a value of $0.05$ in most cases for simulations containing two elements.
This value was reduced to 0.01 for single-component systems.
The DFT parameters employed for each system are briefly discussed in the next section. 
Atomic environments are added to the sparse-set of the GP if their uncertainties are higher than another threshold `sparse-set addition-threshold,' allowing for efficient acquisition of only the most informative atomic environments whose labels are computed with high-fidelity DFT calculations. 
This threshold was set to $20\%$ of the DFT call threshold. 
The relative magnitudes of these thresholds are an important parameter to be tuned, as a smaller sparse-set addition-threshold will add more atomic environments for each DFT call, which could result in the sparse-set of the GP being dominated by non-unique environments and reduce computational efficiency as the hyperparameters are trained during active-learning and this step is dependent on the size of the sparse-set size.
Conversely, setting this value too high (with the practical limit being a $1:1$ match with the DFT threshold) will result in only a few, or even just a single atom, being added to the sparse set for every call to DFT.
To limit the number of duplicate environments added to the sparse-set of the GP from the initial DFT frame, only two atomic environments were manually selected from each surface and NP system, and one atomic environment was selected from bulk or gaseous systems.
The two environments from the surface and NP cells were chosen to reflect both the surface and bulk atomic environments, whereas the selection of a single atomic environment in the bulk- or gaseous-cases was trivial.
Moreover, the atomic positions in the initial frame of each simulation were randomly perturbed by 0.01 \AA{} to push the atomic positions away from their perfect lattice sites.

The FLARE hyperparameters (signal variance $\sigma$, energy noise ($\sigma_E$), force noise ($\sigma_F$), and stress noise ($\sigma_S$)) were optimized during each active learning training simulation up until the 20$^{th}$ DFT call.
Each hyperparameter optimization was allowed to run for a total of $200$ iterations, using the BFGS minimizer, which was found to be sufficiently large for convergence of the hyperparameters.
The priors assigned to each of these hyperparameters were set to empirical values observed previously for bulk Pt FLARE B2 models \cite{owen2023complexity} and from our previous HPt efforts \cite{Vandermause2022,Johansson2022Micron-scaleLearning}, specifically: $3.0$ (eV), $0.001\cdot N_{\text{atoms}}$ (eV), $0.2$ (eV/\AA{}), and $0.001$ (eV/\AA{}$^3$), respectively.
The exact value of these priors is not crucial in the FLARE framework, since hyperparameter optimization is performed and they quickly become dominated by training data once a sufficient number of DFT calls and subsequent optimization steps have been made.

Following the completion of several parallel active learning trajectories, all of the DFT frames were collected, and a final FLARE MLFF was trained on this collection. 
The final training step follows the same selection procedure for atomic environments as described above, where atomic environments are selected based on their relative uncertainties.
This procedure is referred to as `offline-learning,' since the frames have already been calculated at the DFT level and are used only for training the MLFF.
Following creation of the MLFF, we also rescaled the energy noise hyperparameter of the MLFF to account for multiple systems of different sizes and structures being included in the training set.
This hyperparameter describes the noise level of the energy labels of the training dataset and can get trapped in the local minima during the optimization, which affects the model accuracy. 
Therefore, we rescaled the noise hyperparameter of energy labels to 1 meV/atom.

\subsection{Density Functional Theory} \label{sec:dft}
The Vienna \textit{ab initio} Simulation Package (VASP, v5.4.4 \cite{Kresse1993AbMetals,Kresse1996EfficientSet,Kresse1996EfficiencySet,Kresse1999}) was employed for all DFT calculations performed within the FLARE active learning framework and subsequent validation steps. 
All calculations used the generalized gradient approximation (GGA) exchange-correlation functional of Perdew-Burke-Ernzerhof (PBE) \cite{PhysRevLett.77.3865} and projector-augmented wave (PAW) pseudopotentials using the Kresse recommendations in VASP. 
Semi-core corrections of the pseudopotential and spin-polarization were not included for Pt or H. 
A cutoff energy of $450$ eV was employed, with an artificial Methfessel-Paxton temperature \cite{PhysRevB.40.3616} of the electrons set at $0.2$ eV for smearing near the Fermi-energy. 
Brillouin-zone sampling was done on a system-to-system basis, using \textbf{k}-point densities of $0.15$ \AA{}$^{-1}$ for periodic systems and $\Gamma$-point sampling for the NP and gaseous systems and slab systems along the surface normal direction.

\subsection{Production MD Simulations} \label{sec:MD}
The final MD simulations were performed using a custom LAMMPS \cite{THOMPSON2022108171} pairstyle compiled for FLARE \cite{Johansson2022Micron-scaleLearning}.
GPU acceleration was achieved with the Kokkos portability library \cite{CARTEREDWARDS20143202} the performance of which for FLARE has been detailed elsewhere \cite{Johansson2022Micron-scaleLearning}, and is discussed for bulk Pt in the SI.
All simulations for pure Pt employed the Nos\'e-Hoover NVT ensemble with a time-step of $5$ fs, which is appropriate for the mass of Pt, while all simulations containing H employed a time-step of $0.1$ fs.
This latter choice of time-step is more conservative but leads to stable simulations of complex H/Pt environments.
Velocities were randomly initialized for all simulations to a Boltzmann distribution centered at whatever desired temperature for the simulation. 
Each NP was built using the minimized lattice constant of Pt, as predicted by the MLFF and PBE ($3.97$ \AA{}) in the Atomic Simulation Environment (ASE) \cite{Larsen_2017}.
Once each system was equilibrated over the course of $100$ ps, dynamics were then observed at $300$ K for a total of $10$ ns, and thermodynamic and positional information was dumped every $0.1$ ps. 
This dumping frequency was chosen specifically to allow for a mechanistic study of each resulting trajectory. 
For the simulations presented in Fig. \ref{fig:transfer}, the NPs were simulated for a total of 50 ns, and the velocities were initialized to the desired temperature, ranging from 150 to 1500 K.
ASE was also used to construct the H/Pt combined simulations via creation of H$_2$ supercells, which were then added to the NP cells, which were made using the `cluster' class.

\subsection{EXAFS from MD Simulations} \label{sec:mdexafs}
Here we implement a custom code based on Python, OVITO \cite{ovito}, and FEFF10 \cite{kas2021advanced} to run MD-EXAFS calculations for each MLFF trajectory. 
First, the MLFF trajectory, bulk or nano, is surveyed frame-wise to assess MSRD and R versus frame. 
The first frames during any temperature ramping or thermal equilibration are not considered. 
The section of the trajectory at equilibrium, \emph{i.e.}, when MSRD and R versus frame are stabilized, is selected. 
Instead of calculating EXAFS for each frame, we opted to find the minimum number of frames needed to represent the entire trajectory. 
To do this, we sample the MSRD and R versus frame plots with increasingly large frame intervals until MSRD and R versus frame converge. 
For the bulk trajectories, we find that $99$ frames over the equilibrium frames are sufficient, and thus, $99$ frames are sampled per trajectory for MD-EXAFS calculations. 
More frames are needed for the nanoparticles; $250$ frames are sampled per trajectory.

Next, the sampled frames must be readied for EXAFS calculations, and the procedure differs between bulk and nanoparticles, where the former is described in the SI.  
For the nanoparticles, we wanted to calculate EXAFS for every atom in the frame, so procedurally, it was simpler, but the calculation time was much longer. 
For each of the $250$ frames per trajectory, a FEFF input file is generated per atom, resulting in the number of atoms times $250$ per trajectory. 
To speed these calculations up, we employed $8$ working directories with each FEFF calculation using $36$ cores, resulting in approximately $3$ hours per trajectory for $55$ atoms and $12$ hours per trajectory for $147$ atoms (the time scales non-linearly with the number of atoms).

\subsection{Accelerated FEFF Calculations using Neural Networks} \label{sec:nnexafs}
Due to the need for many EXAFS calculations, we developed a surrogate neural network (NN) model to increase computational efficiency. 
To the best of our knowledge, this is the first time a surrogate NN has been used to accelerate FEFF calculations. 
Details of the network training and validation are as follows. 
The method is constructed on the basis that the EXAFS signal is directly related to the radial distribution function (RDF) by equation \ref{eqn:exafs}:

\begin{equation} \label{eqn:exafs}
\chi(k) = S_{0}^{2}\int_{0}^{\infty} g(R)\frac{f(k,R)}{kR^{2}}\sin(2kR+\delta(k))dR
\end{equation}

where \(\chi(k)\) is the EXAFS signal as a function of photoelectron momentum \(k\), \(S_{0}^{2}\) is the amplitude reduction factor, \(g(R)\) is the pair-wise radial distribution function as a function of interatomic distance \(R\) (this equation assumes all pairs are the same element), \(f(k,R)\) is the photoelectron backscattering amplitude, and \(\delta(k)\) is the phase shift. 
Thus, the surrogate NN input is the RDF of the absorbing site of interest determined between 0 and 6 \AA{}, and the output is EXAFS calculated for the site via FEFF10 \cite{kas2021advanced}.

The details of the NN architecture and training are in the SI. 
Briefly, the training data consists of sites taken from the 55 atom and 147 atom cuboctahedral bare and H$_{2}$ exposed NPs and all bulk \textit{fcc} MLFF trajectories. 
The Adam optimizer was used to minimize the MSE loss and care was taken to avoid overfitting. 
In total, 33\% of the absorbing sites were held as a validation set. 
Testing was performed on sites from the 309 atom cuboctahedral bare and H$_{2}$-exposed NP, which the NN had never seen. 
The results of the validation and testing are shown in Fig. S12.

\subsection{Trajectory Analysis} \label{sec:traj}
Trajectory analysis was performed using a custom Python code using the OVITO API and all production MD trajectories were visually inspected with the OVITO software \cite{ovito}. 
First, the coordinates were extracted from MLFF trajectory frames, and for each frame, the Pt atoms were isolated and the Pt-Pt pairwise distances were calculated.
The distances were binned on an r-grid ranging from 0 to 6 \AA{} with a $\Delta$r of 0.025 \AA{} and normalized by $\Delta$r, resulting in the frame-RDF.
For the bulk MLFF trajectories, the Pt atoms are defined by the cluster radius explained in the SI, and for the nanoparticles, every Pt atom is considered.
From the frame-RDF, the mean and variance of the first peak (between 2 and 3.34 \AA{}) were calculated, which corresponded to the R and MSRD, respectively.

\subsection{Temperature Matching} \label{sec:temp}
The experimental to MD temperature conversion is determined by Eqn. \ref{eqn:temp},

\begin{equation} \label{eqn:temp}
T_{\rm MD}(T_{\rm expt})=\frac{{\rm Slope}_{\rm expt}(T_{\rm expt})+Y_{\rm int}^{\rm expt}-Y_{\rm int}^{\rm MD}}{{\rm Slope}_{\rm MD}} 
\end{equation}

where $T_{\textrm{MD}}$ is the MD temperature, $T_{\textrm{expt.}}$ is the experimental temperature, $\textrm{Slope}_{\textrm{expt.}}$, $\textrm{Slope}_{\textrm{MD}}$, $Y_{\rm int}^{\textrm{expt}}$, and $Y_{\rm int}^{\textrm{MD}}$ are the slopes and Y intercepts of the MSRD from MD versus temperature and MSRD from experiment versus temperature plots, respectively. 
Here, the experiment can be defined as the RMC values or the Einstien values. We chose the ideal Einstein values.

\section{Data Availability}
The H-Pt MLFF, all \textit{ab initio} training data, and the training scripts will be provided on the Materials Cloud upon publication.

\subsection*{Author Contributions}
C.J.O.\ augmented the original H/Pt data set from J.V.\ to include NPs using FLARE active learning, performed all MLFF training, validation, MD simulations, and compiled and analyzed the simulation data.
N.M.\ performed EXAFS calculations, EXAFS analysis, RMC calculations, and developed trajectory analysis and MD-EXAFS code. 
C.J.O.\ and N.M.\ jointly created all figures, completed post-processing of the data, and wrote the manuscript. 
Y.X.\ aided in FLARE implementation and calculations, and provided detailed feedback on the FLARE methods section.
A.I.F.\ and R.G.N.\ supervised all aspects of the EXAFS work.
B.K.\ supervised all aspects of the work. 
All authors contributed to revision of the manuscript.

\subsection*{Acknowledgements} 
This work was supported primarily by the US Department of Energy, Office of Basic Energy Sciences Award No. DE-SC0022199 as well as by Robert Bosch LLC. 
C.J.O.\ is supported by the National Science Foundation Graduate Research Fellowship Program under Grant No. DGE1745303. 
This research used resources of the National Energy Research Scientific Computing Center (NERSC), a DOE Office of Science User Facility supported by the Office of Science of the U.S. Department of Energy under Contract No. DE-AC02-05CH11231 using NERSC award BES-ERCAP0024206.
Computing resources were also provided by the FAS Division of Science Research Computing Group at Harvard University, and the Theory and Computation facility of the Center for Functional Nanomaterials (CFN), which is a U.S. Department of Energy Office of Science User Facility, at Brookhaven National Laboratory under Contract No. DE-SC0012704.

\subsection{Competing interests}
The authors declare no competing interests.

\section{References}
\bibliography{bib.bib}

\begin{thebibliography}{10}

\bibitem{Marcella2022}
N.~Marcella \emph{et~al.}, Decoding reactive structures in dilute alloy
  catalysts, {\em Nature Communications}, 13(1), 832, (2022).

\bibitem{Li2021}
Y.~Li \emph{et~al.}, Dynamic structure of active sites in ceria-supported pt
  catalysts for the water gas shift reaction, {\em Nature Communications},
  12(1), 914, (2021).

\bibitem{small_influence_2012}
M.~W. Small, S.~I. Sanchez, N.~S. Marinkovic, A.~I. Frenkel, and R.~G. Nuzzo,
  Influence of {Adsorbates} on the {Electronic} {Structure}, {Bond} {Strain},
  and {Thermal} {Properties} of an {Alumina}-{Supported} {Pt} {Catalyst}, {\em
  ACS Nano}, 6(6), 5583--5595, (2012).

\bibitem{li_noncrystalline--crystalline_2013}
L.~Li \emph{et~al.}, Noncrystalline-to-{Crystalline} {Transformations} in {Pt}
  {Nanoparticles}, {\em Journal of the American Chemical Society}, 135(35),
  13062--13072, (2013).

\bibitem{vila2017anomalous}
F.~D. Vila, J.~J. Rehr, R.~G. Nuzzo, and A.~I. Frenkel, Anomalous structural
  disorder in supported pt nanoparticles, {\em The journal of physical
  chemistry letters}, 8(14), 3284--3288, (2017).

\bibitem{C0CS00089B}
A.~M. Beale, S.~D.~M. Jacques, and B.~M. Weckhuysen, Chemical imaging of
  catalytic solids with synchrotron radiation, {\em Chem. Soc. Rev.}, 39,
  4656--4672, (2010).

\bibitem{doi:10.1021/nl500553a}
H.~L. Xin \emph{et~al.}, Revealing the atomic restructuring of pt–co
  nanoparticles, {\em Nano Letters}, 14(6), 3203--3207, (2014).

\bibitem{doi:10.1021/acs.jpcc.2c05929}
A.~C. Foucher \emph{et~al.}, Atomic-scale stem analysis shows structural
  changes of au–pd nanoparticles in various gaseous environments, {\em The
  Journal of Physical Chemistry C}, 126(42), 18047--18056, (2022).

\bibitem{https://doi.org/10.1002/cctc.201500688}
S.~Zhao \emph{et~al.}, Operando characterization of catalysts through use of a
  portable microreactor, {\em ChemCatChem}, 7(22), 3683--3691, (2015).

\bibitem{Sun2017}
K.~Sun \emph{et~al.}, Operando multi-modal synchrotron investigation for
  structural and chemical evolution of cupric sulfide (cus) additive in li-s
  battery, {\em Scientific Reports}, 7(1), 12976, (2017).

\bibitem{doi:10.1021/acs.jpcc.1c10824}
A.~C. Foucher \emph{et~al.}, Dynamical change of valence states and structure
  in nicu3 nanoparticles during redox cycling, {\em The Journal of Physical
  Chemistry C}, 126(4), 1991--2002, (2022).

\bibitem{doi:10.1021/jacs.2c13666}
A.~C. Foucher \emph{et~al.}, Synthesis and characterization of stable cu–pt
  nanoparticles under reductive and oxidative conditions, {\em Journal of the
  American Chemical Society}, 145(9), 5410--5421, (2023).

\bibitem{doi:10.1021/acscatal.8b01321}
M.~Povia \emph{et~al.}, Combining saxs and xas to study the operando
  degradation of carbon-supported pt-nanoparticle fuel cell catalysts, {\em ACS
  Catalysis}, 8(8), 7000--7015, (2018).

\bibitem{Yan2016}
H.~Yan \emph{et~al.}, Multimodality hard-x-ray imaging of a chromosome with
  nanoscale spatial resolution, {\em Scientific Reports}, 6(1), 20112, (2016).

\bibitem{C4FD00035H}
G.~Smolentsev \emph{et~al.}, X-ray absorption spectroscopy with time-tagged
  photon counting: application to study the structure of a co(i) intermediate
  of h2 evolving photo-catalyst, {\em Faraday Discuss.}, 171, 259--273, (2014).

\bibitem{doi:10.1021/acscatal.2c03863}
R.~Rana, F.~D. Vila, A.~R. Kulkarni, and S.~R. Bare, Bridging the gap between
  the x-ray absorption spectroscopy and the computational catalysis communities
  in heterogeneous catalysis: A perspective on the current and future research
  directions, {\em ACS Catalysis}, 12(22), 13813--13830, (2022).

\bibitem{GREAVES1985203}
G.~Greaves, Exafs and the structure of glass, {\em Journal of Non-Crystalline
  Solids}, 71(1), 203--217, (1985).

\bibitem{Ijima2002}
K.~Ijima, Y.~Ohminami, S.~Suzuki, and K.~Asakura, Polarization-dependent exafs
  measurements of an $\alpha$-molybdenum trioxide single crystal, {\em Topics
  in Catalysis}, 18(1), 125--127, (2002).

\bibitem{PhysRevB.85.195419}
A.~I. Frenkel \emph{et~al.}, Thermal properties of nanoporous gold, {\em Phys.
  Rev. B}, 85, 195419, (2012).

\bibitem{newville2014fundamentals}
M.~Newville, Fundamentals of xafs, {\em Reviews in Mineralogy and
  Geochemistry}, 78(1), 33--74, (2014).

\bibitem{osti_1787395}
M.~Kottwitz, Y.~Li, H.~Wang, A.~I. Frenkel, and R.~G. Nuzzo, Single atom
  catalysts: A review of characterization methods, {\em Chemistry–Methods},
  1(6), (2021).

\bibitem{PhysRevB.48.9825}
E.~A. Stern, Number of relevant independent points in x-ray-absorption
  fine-structure spectra, {\em Phys. Rev. B}, 48, 9825--9827, (1993).

\bibitem{Muller:rv5040}
O.~M{\"{u}}ller, M.~Nachtegaal, J.~Just, D.~L{\"{u}}tzenkirchen-Hecht, and
  R.~Frahm, {Quick-EXAFS setup at the SuperXAS beamline for {\it in~situ} X-ray
  absorption spectroscopy with 10ms time resolution}, {\em Journal of
  Synchrotron Radiation}, 23(1), 260--266, (2016).

\bibitem{kas2021advanced}
J.~Kas, F.~Vila, C.~Pemmaraju, T.~Tan, and J.~Rehr, Advanced calculations of
  x-ray spectroscopies with feff10 and corvus, {\em Journal of Synchrotron
  Radiation}, 28(6), 1801--1810, (2021).

\bibitem{doi:10.1021/jp960160q}
B.~J. Palmer, D.~M. Pfund, and J.~L. Fulton, Direct modeling of exafs spectra
  from molecular dynamics simulations, {\em The Journal of Physical Chemistry},
  100(32), 13393--13398, (1996).

\bibitem{doi:10.1063/1.471711}
P.~D’Angelo, A.~Di~Nola, M.~Mangoni, and N.~V. Pavel, An extended x‐ray
  absorption fine structure study by employing molecular dynamics simulations:
  Bromide ion in methanolic solution, {\em The Journal of Chemical Physics},
  104(5), 1779--1790, (1996).

\bibitem{ANSPOKS20112604}
A.~Anspoks and A.~Kuzmin, Interpretation of the ni k-edge exafs in
  nanocrystalline nickel oxide using molecular dynamics simulations, {\em
  Journal of Non-Crystalline Solids}, 357(14), 2604--2610, (2011).

\bibitem{PhysRevB.85.075439}
S.~W.~T. Price \emph{et~al.}, Fitting exafs data using molecular dynamics
  outputs and a histogram approach, {\em Phys. Rev. B}, 85, 075439, (2012).

\bibitem{Cicco_2002}
A.~D. Cicco \emph{et~al.}, Testing interaction models by using x-ray absorption
  spectroscopy: solid pb, {\em Journal of Physics: Condensed Matter}, 14(12),
  3365, (2002).

\bibitem{PhysRevB.83.115409}
O.~M. Roscioni \emph{et~al.}, Computational prediction of ${L}_{3}$ exafs
  spectra of gold nanoparticles from classical molecular dynamics simulations,
  {\em Phys. Rev. B}, 83, 115409, (2011).

\bibitem{C3SC50614B}
D.~F. Yancey \emph{et~al.}, A theoretical and experimental examination of
  systematic ligand-induced disorder in au dendrimer-encapsulated
  nanoparticles, {\em Chem. Sci.}, 4, 2912--2921, (2013).

\bibitem{doi:10.1021/acsnano.5b00090}
S.~T. Chill \emph{et~al.}, Probing the limits of conventional extended x-ray
  absorption fine structure analysis using thiolated gold nanoparticles, {\em
  ACS Nano}, 9(4), 4036--4042, (2015).

\bibitem{doi:10.1146/annurev-anchem-061318-114929}
J.~Timoshenko, Z.~Duan, G.~Henkelman, R.~Crooks, and A.~Frenkel, Solving the
  structure and dynamics of metal nanoparticles by combining x-ray absorption
  fine structure spectroscopy and atomistic structure simulations, {\em Annual
  Review of Analytical Chemistry}, 12(1), 501--522, (2019).

\bibitem{BOCHAROV2020109198}
D.~Bocharov, M.~Krack, Y.~Rafalskij, A.~Kuzmin, and J.~Purans, Ab initio
  molecular dynamics simulations of negative thermal expansion in scf3: The
  effect of the supercell size, {\em Computational Materials Science}, 171,
  109198, (2020).

\bibitem{SHAPEEV2022111028}
A.~V. Shapeev, D.~Bocharov, and A.~Kuzmin, Validation of moment tensor
  potentials for fcc and bcc metals using exafs spectra, {\em Computational
  Materials Science}, 210, 111028, (2022).

\bibitem{doi:10.1021/acs.jpclett.1c01204}
I.~Poltavsky and A.~Tkatchenko, Machine learning force fields: Recent advances
  and remaining challenges, {\em The Journal of Physical Chemistry Letters},
  12(28), 6551--6564, (2021).

\bibitem{https://doi.org/10.1002/adma.201902765}
V.~L. Deringer, M.~A. Caro, and G.~Csányi, Machine learning interatomic
  potentials as emerging tools for materials science, {\em Advanced Materials},
  31(46), 1902765, (2019).

\bibitem{Vandermause2022}
J.~Vandermause, Y.~Xie, J.~S. Lim, C.~J. Owen, and B.~Kozinsky, Active learning
  of reactive bayesian force fields applied to heterogeneous catalysis dynamics
  of h/pt, {\em Nature Communications}, 13(1), 5183, (2022).

\bibitem{Lim2020EvolutionDynamics}
J.~S. Lim \emph{et~al.}, {Evolution of Metastable Structures at Bimetallic
  Surfaces from Microscopy and Machine-Learning Molecular Dynamics}, {\em
  Journal of the American Chemical Society}, 142(37), 15907--15916, (2020).

\bibitem{Kovacs2021LinearRMSE}
D.~P. Kov{\'{a}}cs \emph{et~al.}, {Linear Atomic Cluster Expansion Force Fields
  for Organic Molecules: Beyond RMSE}, {\em Journal of Chemical Theory and
  Computation}, 17(12), 7696--7711, (2021).

\bibitem{Batzner2021E3-EquivariantPotentials}
S.~Batzner \emph{et~al.}, {E(3)-Equivariant Graph Neural Networks for
  Data-Efficient and Accurate Interatomic Potentials}, {\em Nature
  Communications 2022}, 13(1), 1--11, (2021).

\bibitem{Musaelian2023}
A.~Musaelian \emph{et~al.}, Learning local equivariant representations for
  large-scale atomistic dynamics, {\em Nature Communications}, 14(1), 579,
  (2023).

\bibitem{owen2023complexity}
C.~J. Owen \emph{et~al.}, Complexity of many-body interactions in transition
  metals via machine-learned force fields from the tm23 data set, {\em arXiv
  preprint arXiv:2302.12993}, (2023).

\bibitem{Vandermause2020}
J.~Vandermause \emph{et~al.}, On-the-fly active learning of interpretable
  bayesian force fields for atomistic rare events, {\em npj Computational
  Materials}, 6(1), 20, (2020).

\bibitem{https://doi.org/10.48550/arxiv.2211.09866}
A.~Zhu, S.~Batzner, A.~Musaelian, and B.~Kozinsky, Fast uncertainty estimates
  in deep learning interatomic potentials, {\em arXiv preprint
  arXiv:2211.09866}, (2022).

\bibitem{Johansson2022Micron-scaleLearning}
A.~Johansson \emph{et~al.}, {Micron-scale heterogeneous catalysis with Bayesian
  force fields from first principles and active learning}, {\em arXiv preprint
  arXiv:2204.12573}, (2022).

\bibitem{PhysRevB.81.115451}
A.~Yevick and A.~I. Frenkel, Effects of surface disorder on exafs modeling of
  metallic clusters, {\em Phys. Rev. B}, 81, 115451, (2010).

\bibitem{frenkel_view_2001}
A.~I. Frenkel, C.~W. Hills, and R.~G. Nuzzo, A {View} from the {Inside}:
  {Complexity} in the {Atomic} {Scale} {Ordering} of {Supported} {Metal}
  {Nanoparticles}, {\em The Journal of Physical Chemistry B}, 105(51),
  12689--12703, (2001).

\bibitem{rodriguez2013situ}
J.~A. Rodr{\'\i}guez, J.~C. Hanson, and P.~J. Chupas.
\newblock {\em In-situ characterization of heterogeneous catalysts}.
\newblock John Wiley \& Sons, Ltd, (2013).

\bibitem{doi:10.1021/acs.jpcc.3c00571}
P.~K. Routh, N.~Marcella, and A.~I. Frenkel, Speciation of nanocatalysts using
  x-ray absorption spectroscopy assisted by machine learning, {\em The Journal
  of Physical Chemistry C}, 127(12), 5653--5662, (2023).

\bibitem{phan_what_2019}
H.~T. Phan and A.~J. Haes, What {Does} {Nanoparticle} {Stability} {Mean}?, {\em
  The Journal of Physical Chemistry C}, 123(27), 16495--16507, (2019).

\bibitem{SanchezNonBulk}
S.~I. Sanchez \emph{et~al.}, The emergence of nonbulk properties in supported
  metal clusters: Negative thermal expansion and atomic disorder in pt
  nanoclusters supported on $\gamma$-al2o3, {\em Journal of the American
  Chemical Society}, 131(20), 7040--7054, (2009).

\bibitem{Timoshenko2017b}
J.~Timoshenko, D.~Lu, Y.~Lin, and A.~I. Frenkel, {Supervised
  Machine-Learning-Based Determination of Three-Dimensional Structure of
  Metallic Nanoparticles}, {\em Journal of Physical Chemistry Letters}, 8(20),
  5091--5098, (2017).

\bibitem{Frenkel2012}
A.~I. Frenkel, J.~A. Rodriguez, and J.~G. Chen, {Synchrotron techniques for in
  situ catalytic studies: Capabilities, challenges, and opportunities}, {\em
  ACS Catalysis}, 2(11), 2269--2280, (2012).

\bibitem{RevModPhys.72.621}
J.~J. Rehr and R.~C. Albers, Theoretical approaches to x-ray absorption fine
  structure, {\em Rev. Mod. Phys.}, 72, 621--654, (2000).

\bibitem{Larsen_2017}
A.~H. Larsen \emph{et~al.}, The atomic simulation environment—a python
  library for working with atoms, {\em Journal of Physics: Condensed Matter},
  29(27), 273002, (2017).

\bibitem{frenkel2014critical}
A.~I. Frenkel \emph{et~al.}, Critical review: effects of complex interactions
  on structure and dynamics of supported metal catalysts, {\em Journal of
  Vacuum Science \& Technology A: Vacuum, Surfaces, and Films}, 32(2), 020801,
  (2014).

\bibitem{frenkel2013situ}
A.~I. Frenkel \emph{et~al.}, An in situ study of bond strains in 1 nm pt
  catalysts and their sensitivities to cluster--support and cluster--adsorbate
  interactions, {\em The Journal of Physical Chemistry C}, 117(44),
  23286--23294, (2013).

\bibitem{Surnev_2001}
S.~Surnev, M.~G. Ramsey, and F.~P. Netzer, Synchrotron radiation applied to the
  study of heterogeneous model catalyst surfaces, {\em Journal of Physics:
  Condensed Matter}, 13(49), 11305, (2001).

\bibitem{doi:10.1021/cs3004006}
A.~I. Frenkel, J.~A. Rodriguez, and J.~G. Chen, Synchrotron techniques for in
  situ catalytic studies: Capabilities, challenges, and opportunities, {\em ACS
  Catalysis}, 2(11), 2269--2280, (2012).

\bibitem{catal8100450}
C.~Yu \emph{et~al.}, H2 thermal desorption spectra on pt(111): A density
  functional theory and kinetic monte carlo simulation study, {\em Catalysts},
  8(10), (2018).

\bibitem{doi:10.1098/rsta.1991.0021}
V.~Heine \emph{et~al.}, Many-atom interactions in solids, {\em Philosophical
  Transactions of the Royal Society of London. Series A: Physical and
  Engineering Sciences}, 334(1635), 393--405, (1991).

\bibitem{doi:https://doi.org/10.1002/9781118844243.ch3}
J.~J. Kas, K.~Jorissen, and J.~J. Rehr.
\newblock {\em Real-Space Multiple-Scattering Theory of X-Ray Spectra},
  chapter~3, pp. 51--72.
\newblock John Wiley {\&} Sons, Ltd, (2016).

\bibitem{Larsen_2016}
P.~M. Larsen, S.~Schmidt, and J.~Schiøtz, Robust structural identification via
  polyhedral template matching, {\em Modelling and Simulation in Materials
  Science and Engineering}, 24(5), 055007, (2016).

\bibitem{ovito}
A.~Stukowski, {Visualization and analysis of atomistic simulation data with
  OVITO-the Open Visualization Tool}, {\em {Modelling and Simulation in
  Materials Science and Engineering}}, {18}({1}), ({2010}).

\bibitem{PhysRevLett.93.065502}
E.~Apr\`a, F.~Baletto, R.~Ferrando, and A.~Fortunelli, Amorphization mechanism
  of icosahedral metal nanoclusters, {\em Phys. Rev. Lett.}, 93, 065502,
  (2004).

\bibitem{MILLER1993395}
J.~Miller \emph{et~al.}, Hydrogen temperature-programmed desorption (h2 tpd) of
  supported platinum catalysts, {\em Journal of Catalysis}, 143(2), 395--408,
  (1993).

\bibitem{Drautz2019AtomicPotentials}
R.~Drautz, {Atomic cluster expansion for accurate and transferable interatomic
  potentials}, {\em Physical Review B}, 99(1), 014104, (2019).

\bibitem{Xie2021BayesianStanene}
Y.~Xie, J.~Vandermause, L.~Sun, A.~Cepellotti, and B.~Kozinsky, {Bayesian force
  fields from active learning for simulation of inter-dimensional
  transformation of stanene}, {\em npj Computational Materials}, 7(1), 1--10,
  (2021).

\bibitem{Xie2023Uncertainty-awareSiC}
Y.~Xie \emph{et~al.}, Uncertainty-aware molecular dynamics from bayesian active
  learning for phase transformations and thermal transport in sic, {\em npj
  Computational Materials}, 9(1), 36, (2023).

\bibitem{Kresse1993AbMetals}
G.~Kresse and J.~Hafner, {Ab initio molecular dynamics for liquid metals}, {\em
  Physical Review B}, 47(1), 558--561, (1993).

\bibitem{Kresse1996EfficientSet}
G.~Kresse and J.~Furthm{\"{u}}ller, {Efficient iterative schemes for ab initio
  total-energy calculations using a plane-wave basis set}, {\em Physical Review
  B - Condensed Matter and Materials Physics}, 54(16), 11169--11186, (1996).

\bibitem{Kresse1996EfficiencySet}
G.~Kresse and J.~Furthm{\"{u}}ller, {Efficiency of ab-initio total energy
  calculations for metals and semiconductors using a plane-wave basis set},
  {\em Computational Materials Science}, 6(1), 15--50, (1996).

\bibitem{Kresse1999}
G.~Kresse and D.~Joubert, {From ultrasoft pseudopotentials to the projector
  augmented-wave method}, {\em Physical Review B - Condensed Matter and
  Materials Physics}, 59(3), 1758--1775, (1999).

\bibitem{PhysRevLett.77.3865}
J.~P. Perdew, K.~Burke, and M.~Ernzerhof, Generalized gradient approximation
  made simple, {\em Phys. Rev. Lett.}, 77, 3865--3868, (1996).

\bibitem{PhysRevB.40.3616}
M.~Methfessel and A.~T. Paxton, High-precision sampling for brillouin-zone
  integration in metals, {\em Phys. Rev. B}, 40, 3616--3621, (1989).

\bibitem{THOMPSON2022108171}
A.~P. Thompson \emph{et~al.}, Lammps - a flexible simulation tool for
  particle-based materials modeling at the atomic, meso, and continuum scales,
  {\em Computer Physics Communications}, 271, 108171, (2022).

\bibitem{CARTEREDWARDS20143202}
H.~{Carter Edwards}, C.~R. Trott, and D.~Sunderland, Kokkos: Enabling manycore
  performance portability through polymorphic memory access patterns, {\em
  Journal of Parallel and Distributed Computing}, 74(12), 3202--3216, (2014).

\end{thebibliography}


\begin{thebibliography}{10}

\bibitem{timoshenko2014exafs}
J.~Timoshenko, A.~Kuzmin, and J.~Purans, Exafs study of hydrogen intercalation
  into reo 3 using the evolutionary algorithm, {\em Journal of Physics:
  Condensed Matter}, 26(5), 055401, (2014).

\bibitem{Rehr2010Parameter-freeFEFF9}
J.~J. Rehr, J.~J. Kas, F.~D. Vila, M.~P. Prange, and K.~Jorissen.
\newblock {Parameter-free calculations of X-ray spectra with FEFF9}, (2010).

\bibitem{frenkel_view_2001}
A.~I. Frenkel, C.~W. Hills, and R.~G. Nuzzo, A {View} from the {Inside}:
  {Complexity} in the {Atomic} {Scale} {Ordering} of {Supported} {Metal}
  {Nanoparticles}, {\em The Journal of Physical Chemistry B}, 105(51),
  12689--12703, (2001).

\bibitem{Ravel:ph5155}
B.~Ravel and M.~Newville, {{\it ATHENA}, {\it ARTEMIS}, {\it HEPHAESTUS}: data
  analysis for X-ray absorption spectroscopy using {\it IFEFFIT}}, {\em Journal
  of Synchrotron Radiation}, 12(4), 537--541, (2005).

\bibitem{owen2023complexity}
C.~J. Owen \emph{et~al.}, Complexity of many-body interactions in transition
  metals via machine-learned force fields from the tm23 data set, {\em arXiv
  preprint arXiv:2302.12993}, (2023).

\bibitem{SanchezNonBulk}
S.~I. Sanchez \emph{et~al.}, The emergence of nonbulk properties in supported
  metal clusters: Negative thermal expansion and atomic disorder in pt
  nanoclusters supported on $\gamma$-al2o3, {\em Journal of the American
  Chemical Society}, 131(20), 7040--7054, (2009).

\bibitem{PhysRevB.48.585}
A.~I. Frenkel and J.~J. Rehr, Thermal expansion and x-ray-absorption
  fine-structure cumulants, {\em Phys. Rev. B}, 48, 585--588, (1993).

\bibitem{kas2021advanced}
J.~Kas, F.~Vila, C.~Pemmaraju, T.~Tan, and J.~Rehr, Advanced calculations of
  x-ray spectroscopies with feff10 and corvus, {\em Journal of Synchrotron
  Radiation}, 28(6), 1801--1810, (2021).

\bibitem{PhysRevB.20.4908}
E.~Sevillano, H.~Meuth, and J.~J. Rehr, Extended x-ray absorption fine
  structure debye-waller factors. i. monatomic crystals, {\em Phys. Rev. B},
  20, 4908--4911, (1979).

\bibitem{WELLENDORFF201536}
J.~Wellendorff \emph{et~al.}, A benchmark database for adsorption bond energies
  to transition metal surfaces and comparison to selected dft functionals, {\em
  Surface Science}, 640, 36--44, (2015).
\newblock Reactivity Concepts at Surfaces: Coupling Theory with Experiment.

\bibitem{Vandermause2022}
J.~Vandermause, Y.~Xie, J.~S. Lim, C.~J. Owen, and B.~Kozinsky, Active learning
  of reactive bayesian force fields applied to heterogeneous catalysis dynamics
  of h/pt, {\em Nature Communications}, 13(1), 5183, (2022).

\bibitem{Johansson2022Micron-scaleLearning}
A.~Johansson \emph{et~al.}, {Micron-scale heterogeneous catalysis with Bayesian
  force fields from first principles and active learning}, {\em arXiv preprint
  arXiv:2204.12573}, (2022).

\bibitem{doi:10.1021/ja077011d}
Crc handbook of chemistry and physics, 88th ed editor-in-chief: David r. lide
  (national institute of standards and technology) crc press/taylor \& francis
  group: Boca raton, fl. 2007. 2640 pp. \$139.95. isbn 0-8493-0488-1., {\em
  Journal of the American Chemical Society}, 130(1), 382--382, (2008).

\end{thebibliography}



ENTRY
  { address
    author
    booktitle
    chapter
    edition
    editor
    howpublished
    institution
    journal
    key
    month
    note
    number
    organization
    pages
    publisher
    school
    series
    title
    type
    volume
    year
  }
  {}
  { label }

INTEGERS { output.state before.all mid.sentence after.sentence after.block }

FUNCTION {init.state.consts}
{ #0 'before.all :=
  #1 'mid.sentence :=
  #2 'after.sentence :=
  #3 'after.block :=
}

STRINGS { s t }

FUNCTION {output.nonnull}
{ 's :=
  output.state mid.sentence =
    { ", " * write$ }
    { output.state after.block =
        { add.period$ write$
          newline$
          "\newblock " write$
        }
        { output.state before.all =
            'write$
            { add.period$ " " * write$ }
          if$
        }
      if$
      mid.sentence 'output.state :=
    }
  if$
  s
}

FUNCTION {output}
{ duplicate$ empty$
    'pop$
    'output.nonnull
  if$
}

FUNCTION {output.check}
{ 't :=
  duplicate$ empty$
    { pop$ "empty " t * " in " * cite$ * warning$ }
    'output.nonnull
  if$
}

FUNCTION {output.bibitem}
{ newline$
  "\bibitem{" write$
  cite$ write$
  "}" write$
  newline$
  ""
  before.all 'output.state :=
}

FUNCTION {fin.entry}
{ add.period$
  write$
  newline$
}

FUNCTION {new.block}
{ output.state before.all =
    'skip$
    { after.block 'output.state := }
  if$
}

FUNCTION {new.sentence}
{ output.state after.block =
    'skip$
    { output.state before.all =
        'skip$
        { after.sentence 'output.state := }
      if$
    }
  if$
}

FUNCTION {not}
{   { #0 }
    { #1 }
  if$
}

FUNCTION {and}
{   'skip$
    { pop$ #0 }
  if$
}

FUNCTION {or}
{   { pop$ #1 }
    'skip$
  if$
}

FUNCTION {new.block.checka}
{ empty$
    'skip$
    'new.block
  if$
}

FUNCTION {new.block.checkb}
{ empty$
  swap$ empty$
  and
    'skip$
    'new.block
  if$
}

FUNCTION {new.sentence.checka}
{ empty$
    'skip$
    'new.sentence
  if$
}

FUNCTION {new.sentence.checkb}
{ empty$
  swap$ empty$
  and
    'skip$
    'new.sentence
  if$
}

FUNCTION {field.or.null}
{ duplicate$ empty$
    { pop$ "" }
    'skip$
  if$
}

FUNCTION {emphasize}
{ duplicate$ empty$
    { pop$ "" }
    { "{\em " swap$ * "}" * }
  if$
}

INTEGERS { nameptr namesleft numnames }


FUNCTION {format.names}
{ 's :=
  #1 'nameptr :=
  s num.names$ 'numnames :=
  numnames 'namesleft :=
  numnames #5 >
    { s #1 "{f.~}{vv~}{ll}{, jj}" format.name$
      " \emph{et~al.}" * }
    {
      { namesleft #0 > }
      { s nameptr "{f.~}{vv~}{ll}{, jj}" format.name$ 't :=
        nameptr #1 >
          { namesleft #1 >
              { ", " * t * }
              { numnames #2 >
                  { "," * }
                  'skip$
                if$
                t "others" =
                  { " \emph{et~al}." * }
                  { " and " * t * }
                if$
              }
            if$
          }
          't
        if$
        nameptr #1 + 'nameptr :=
        namesleft #1 - 'namesleft :=
      }
    while$
  }
  if$
}


FUNCTION {format.authors}
{ author empty$
    { "" }
    { author format.names }
  if$
}

FUNCTION {format.editors}
{ editor empty$
    { "" }
    { editor format.names
      editor num.names$ #1 >
        { ", editors" * }
        { ", editor" * }
      if$
    }
  if$
}

FUNCTION {format.title}
{ title empty$
    { "" }
    { title "t" change.case$ }
  if$
}

FUNCTION {n.dashify}
{ 't :=
  ""
    { t empty$ not }
    { t #1 #1 substring$ "-" =
        { t #1 #2 substring$ "--" = not
            { "--" *
              t #2 global.max$ substring$ 't :=
            }
            {   { t #1 #1 substring$ "-" = }
                { "-" *
                  t #2 global.max$ substring$ 't :=
                }
              while$
            }
          if$
        }
        { t #1 #1 substring$ *
          t #2 global.max$ substring$ 't :=
        }
      if$
    }
  while$
}

FUNCTION {format.date}
{ "("  year ")" * *
}

FUNCTION {format.btitle}
{ title emphasize
}

FUNCTION {tie.or.space.connect}
{ duplicate$ text.length$ #3 <
    { "~" }
    { " " }
  if$
  swap$ * *
}

FUNCTION {either.or.check}
{ empty$
    'pop$
    { "can't use both " swap$ * " fields in " * cite$ * warning$ }
  if$
}

FUNCTION {format.bvolume}
{ volume empty$
    { "" }
    { "volume" volume tie.or.space.connect
      series empty$
        'skip$
        { " of " * series emphasize * }
      if$
      "volume and number" number either.or.check
    }
  if$
}

FUNCTION {format.number.series}
{ volume empty$
    { number empty$
        { series field.or.null }
        { output.state mid.sentence =
            { "number" }
            { "Number" }
          if$
          number tie.or.space.connect
          series empty$
            { "there's a number but no series in " cite$ * warning$ }
            { " in " * series * }
          if$
        }
      if$
    }
    { "" }
  if$
}

FUNCTION {format.edition}
{ edition empty$
    { "" }
    { output.state mid.sentence =
        { edition "l" change.case$ " edition" * }
        { edition "t" change.case$ " edition" * }
      if$
    }
  if$
}

INTEGERS { multiresult }

FUNCTION {multi.page.check}
{ 't :=
  #0 'multiresult :=
    { multiresult not
      t empty$ not
      and
    }
    { t #1 #1 substring$
      duplicate$ "-" =
      swap$ duplicate$ "," =
      swap$ "+" =
      or or
        { #1 'multiresult := }
        { t #2 global.max$ substring$ 't := }
      if$
    }
  while$
  multiresult
}

FUNCTION {format.pages}
{ pages empty$
    { "" }
    { pages multi.page.check
        { "pp." pages n.dashify tie.or.space.connect }
        { "pp." pages tie.or.space.connect }
      if$
    }
  if$
}

FUNCTION {format.vol.num.pages}
{ volume field.or.null
  number empty$
    'skip$
    { "(" number * ")" * *
      volume empty$
        { "there's a number but no volume in " cite$ * warning$ }
        'skip$
      if$
    }
  if$
  pages empty$
    'skip$
    { duplicate$ empty$
        { pop$ format.pages }
        { ", " * pages n.dashify * }
      if$
    }
  if$
}

FUNCTION {format.chapter.pages}
{ chapter empty$
    'format.pages
    { type empty$
        { "chapter" }
        { type "l" change.case$ }
      if$
      chapter tie.or.space.connect
      pages empty$
        'skip$
        { ", " * format.pages * }
      if$
    }
  if$
}

FUNCTION {format.in.ed.booktitle}
{ booktitle empty$
    { "" }
    { editor empty$
        { "In " booktitle emphasize * }
        { "In " format.editors * ", " * booktitle emphasize * }
      if$
    }
  if$
}

FUNCTION {empty.misc.check}
{ author empty$ title empty$ howpublished empty$
  month empty$ year empty$ note empty$
  and and and and and
    { "all relevant fields are empty in " cite$ * warning$ }
    'skip$
  if$
}

FUNCTION {format.thesis.type}
{ type empty$
    'skip$
    { pop$
      type "t" change.case$
    }
  if$
}

FUNCTION {format.tr.number}
{ type empty$
    { "Technical Report" }
    'type
  if$
  number empty$
    { "t" change.case$ }
    { number tie.or.space.connect }
  if$
}

FUNCTION {format.article.crossref}
{ key empty$
    { journal empty$
        { "need key or journal for " cite$ * " to crossref " * crossref *
          warning$
          ""
        }
        { "In {\em " journal * "\/}" * }
      if$
    }
    { "In " key * }
  if$
  " \cite{" * crossref * "}" *
}

FUNCTION {format.crossref.editor}
{ editor #1 "{vv~}{ll}" format.name$
  editor num.names$ duplicate$
  #2 >
    { pop$ " et~al." * }
    { #2 <
        'skip$
        { editor #2 "{ff }{vv }{ll}{ jj}" format.name$ "others" =
            { " et~al." * }
            { " and " * editor #2 "{vv~}{ll}" format.name$ * }
          if$
        }
      if$
    }
  if$
}

FUNCTION {format.book.crossref}
{ volume empty$
    { "empty volume in " cite$ * "'s crossref of " * crossref * warning$
      "In "
    }
    { "Volume" volume tie.or.space.connect
      " of " *
    }
  if$
  editor empty$
  editor field.or.null author field.or.null =
  or
    { key empty$
        { series empty$
            { "need editor, key, or series for " cite$ * " to crossref " *
              crossref * warning$
              "" *
            }
            { "{\em " * series * "\/}" * }
          if$
        }
        { key * }
      if$
    }
    { format.crossref.editor * }
  if$
  " \cite{" * crossref * "}" *
}

FUNCTION {format.incoll.inproc.crossref}
{ editor empty$
  editor field.or.null author field.or.null =
  or
    { key empty$
        { booktitle empty$
            { "need editor, key, or booktitle for " cite$ * " to crossref " *
              crossref * warning$
              ""
            }
            { "In {\em " booktitle * "\/}" * }
          if$
        }
        { "In " key * }
      if$
    }
    { "In " format.crossref.editor * }
  if$
  " \cite{" * crossref * "}" *
}

FUNCTION {article}
{ output.bibitem
  format.authors "author" output.check
  format.title "title" output.check
  crossref missing$
    { journal emphasize "journal" output.check
      format.vol.num.pages output
      format.date "year" output.check
    }
    { format.article.crossref output.nonnull
      format.pages output
    }
  if$
  new.block
  note output
  fin.entry
}

FUNCTION {book}
{ output.bibitem
  author empty$
    { format.editors "author and editor" output.check }
    { format.authors output.nonnull
      crossref missing$
        { "author and editor" editor either.or.check }
        'skip$
      if$
    }
  if$
  new.block
  format.btitle "title" output.check
  crossref missing$
    { format.bvolume output
      new.block
      format.number.series output
      new.sentence
      publisher "publisher" output.check
      address output
    }
    { new.block
      format.book.crossref output.nonnull
    }
  if$
  format.edition output
  format.date "year" output.check
  new.block
  note output
  fin.entry
}

FUNCTION {booklet}
{ output.bibitem
  format.authors output
  new.block
  format.title "title" output.check
  howpublished address new.block.checkb
  howpublished output
  address output
  format.date output
  new.block
  note output
  fin.entry
}

FUNCTION {inbook}
{ output.bibitem
  author empty$
    { format.editors "author and editor" output.check }
    { format.authors output.nonnull
      crossref missing$
        { "author and editor" editor either.or.check }
        'skip$
      if$
    }
  if$
  new.block
  format.btitle "title" output.check
  crossref missing$
    { format.bvolume output
      format.chapter.pages "chapter and pages" output.check
      new.block
      format.number.series output
      new.sentence
      publisher "publisher" output.check
      address output
    }
    { format.chapter.pages "chapter and pages" output.check
      new.block
      format.book.crossref output.nonnull
    }
  if$
  format.edition output
  format.date "year" output.check
  new.block
  note output
  fin.entry
}

FUNCTION {incollection}
{ output.bibitem
  format.authors "author" output.check
  new.block
  format.title "title" output.check
  new.block
  crossref missing$
    { format.in.ed.booktitle "booktitle" output.check
      format.bvolume output
      format.number.series output
      format.chapter.pages output
      new.sentence
      publisher "publisher" output.check
      address output
      format.edition output
      format.date "year" output.check
    }
    { format.incoll.inproc.crossref output.nonnull
      format.chapter.pages output
    }
  if$
  new.block
  note output
  fin.entry
}

FUNCTION {inproceedings}
{ output.bibitem
  format.authors "author" output.check
  new.block
  format.title "title" output.check
  new.block
  crossref missing$
    { format.in.ed.booktitle "booktitle" output.check
      format.bvolume output
      format.number.series output
      format.pages output
      address empty$
        { organization publisher new.sentence.checkb
          organization output
          publisher output
          format.date "year" output.check
        }
        { address output.nonnull
          format.date "year" output.check
          new.sentence
          organization output
          publisher output
        }
      if$
    }
    { format.incoll.inproc.crossref output.nonnull
      format.pages output
    }
  if$
  new.block
  note output
  fin.entry
}

FUNCTION {conference} { inproceedings }

FUNCTION {manual}
{ output.bibitem
  author empty$
    { organization empty$
        'skip$
        { organization output.nonnull
          address output
        }
      if$
    }
    { format.authors output.nonnull }
  if$
  new.block
  format.btitle "title" output.check
  author empty$
    { organization empty$
        { address new.block.checka
          address output
        }
        'skip$
      if$
    }
    { organization address new.block.checkb
      organization output
      address output
    }
  if$
  format.edition output
  format.date output
  new.block
  note output
  fin.entry
}

FUNCTION {mastersthesis}
{ output.bibitem
  format.authors "author" output.check
  new.block
  format.title "title" output.check
  new.block
  "Master's thesis" format.thesis.type output.nonnull
  school "school" output.check
  address output
  format.date "year" output.check
  new.block
  note output
  fin.entry
}

FUNCTION {misc}
{ output.bibitem
  format.authors output
  title howpublished new.block.checkb
  format.title output
  howpublished new.block.checka
  howpublished output
  format.date output
  new.block
  note output
  fin.entry
  empty.misc.check
}

FUNCTION {phdthesis}
{ output.bibitem
  format.authors "author" output.check
  new.block
  format.btitle "title" output.check
  new.block
  "PhD thesis" format.thesis.type output.nonnull
  school "school" output.check
  address output
  format.date "year" output.check
  new.block
  note output
  fin.entry
}

FUNCTION {proceedings}
{ output.bibitem
  editor empty$
    { organization output }
    { format.editors output.nonnull }
  if$
  new.block
  format.btitle "title" output.check
  format.bvolume output
  format.number.series output
  address empty$
    { editor empty$
        { publisher new.sentence.checka }
        { organization publisher new.sentence.checkb
          organization output
        }
      if$
      publisher output
      format.date "year" output.check
    }
    { address output.nonnull
      format.date "year" output.check
      new.sentence
      editor empty$
        'skip$
        { organization output }
      if$
      publisher output
    }
  if$
  new.block
  note output
  fin.entry
}

FUNCTION {techreport}
{ output.bibitem
  format.authors "author" output.check
  new.block
  format.title "title" output.check
  new.block
  format.tr.number output.nonnull
  institution "institution" output.check
  address output
  format.date "year" output.check
  new.block
  note output
  fin.entry
}

FUNCTION {unpublished}
{ output.bibitem
  format.authors "author" output.check
  new.block
  format.title "title" output.check
  new.block
  note "note" output.check
  format.date output
  fin.entry
}

FUNCTION {default.type} { misc }

MACRO {jan} {"January"}

MACRO {feb} {"February"}

MACRO {mar} {"March"}

MACRO {apr} {"April"}

MACRO {may} {"May"}

MACRO {jun} {"June"}

MACRO {jul} {"July"}

MACRO {aug} {"August"}

MACRO {sep} {"September"}

MACRO {oct} {"October"}

MACRO {nov} {"November"}

MACRO {dec} {"December"}

MACRO {acmcs} {"ACM Computing Surveys"}

MACRO {acta} {"Acta Informatica"}

MACRO {cacm} {"Communications of the ACM"}

MACRO {ibmjrd} {"IBM Journal of Research and Development"}

MACRO {ibmsj} {"IBM Systems Journal"}

MACRO {ieeese} {"IEEE Transactions on Software Engineering"}

MACRO {ieeetc} {"IEEE Transactions on Computers"}

MACRO {ieeetcad}
 {"IEEE Transactions on Computer-Aided Design of Integrated Circuits"}

MACRO {ipl} {"Information Processing Letters"}

MACRO {jacm} {"Journal of the ACM"}

MACRO {jcss} {"Journal of Computer and System Sciences"}

MACRO {scp} {"Science of Computer Programming"}

MACRO {sicomp} {"SIAM Journal on Computing"}

MACRO {tocs} {"ACM Transactions on Computer Systems"}

MACRO {tods} {"ACM Transactions on Database Systems"}

MACRO {tog} {"ACM Transactions on Graphics"}

MACRO {toms} {"ACM Transactions on Mathematical Software"}

MACRO {toois} {"ACM Transactions on Office Information Systems"}

MACRO {toplas} {"ACM Transactions on Programming Languages and Systems"}

MACRO {tcs} {"Theoretical Computer Science"}

READ

STRINGS { longest.label }

INTEGERS { number.label longest.label.width }

FUNCTION {initialize.longest.label}
{ "" 'longest.label :=
  #1 'number.label :=
  #0 'longest.label.width :=
}

FUNCTION {longest.label.pass}
{ number.label int.to.str$ 'label :=
  number.label #1 + 'number.label :=
  label width$ longest.label.width >
    { label 'longest.label :=
      label width$ 'longest.label.width :=
    }
    'skip$
  if$
}

EXECUTE {initialize.longest.label}

ITERATE {longest.label.pass}

FUNCTION {begin.bib}
{ preamble$ empty$
    'skip$
    { preamble$ write$ newline$ }
  if$
  "\begin{thebibliography}{"  longest.label  * "}" * write$ newline$
}

EXECUTE {begin.bib}

EXECUTE {init.state.consts}

ITERATE {call.type$}

FUNCTION {end.bib}
{ newline$
  "\end{thebibliography}" write$ newline$
}

EXECUTE {end.bib}

\end{document}


\title{Supplementary Information for: Unraveling the Catalytic Effect of Hydrogen Adsorption on Pt Nanoparticle Shape-Change}

\author{Cameron J. Owen$^{*,\dagger}$}
\affiliation{Department of Chemistry and Chemical Biology, Harvard University, Cambridge, Massachusetts 02138, United States}

\author{Nicholas Marcella$^{*,\dagger}$}
\affiliation{Department of Chemistry, University of Illinois, Urbana, Illinois 61801, United States}

\author{\\Yu Xie}
\affiliation{John A. Paulson School of Engineering and Applied Sciences, Harvard University, Cambridge, Massachusetts 02138, United States}

\author{Jonathan Vandermause$^{\ddagger}$}
\affiliation{John A. Paulson School of Engineering and Applied Sciences, Harvard University, Cambridge, Massachusetts 02138, United States}

\author{Anatoly I. Frenkel}
\affiliation{Department of Materials Science and Chemical Engineering, Stony Brook University, Stony Brook, New York 11794, United States}
\affiliation{Chemistry Division, Brookhaven National Laboratory, Upton, New York 11973, United States}

\author{Ralph G. Nuzzo}
\affiliation{Department of Chemistry, University of Illinois, Urbana, Illinois 61801, United States}

\author{Boris Kozinsky$^{\dagger,}$}
\affiliation{John A. Paulson School of Engineering and Applied Sciences, Harvard University, Cambridge, Massachusetts 02138, United States}
\affiliation{Robert Bosch LLC Research and Technology Center, Cambridge, Massachusetts 02139, United States}

\def\thefootnote{$*$}\footnotetext{These authors contributed equally.}\def\thefootnote{\arabic{footnote}}

\def\thefootnote{$\ddagger$}\footnotetext{Currently at D.E. Shaw Research.}\def\thefootnote{\arabic{footnote}}

\def\thefootnote{$\dagger$}\footnotetext{Corresponding authors\\C.J.O., E-mail: \url{cowen@g.harvard.edu}\\N.M., E-mail: \url{nmarcella@bnl.gov}\\B.K., E-mail: \url{bkoz@seas.harvard.edu}\def\thefootnote{\arabic{footnote}}}

\newcommand\bvec{\mathbf}
\newcommand{\mathsc}[1]{{\normalfont\textsc{#1}}}

\maketitle

\section{Supplementary Methods}
\subsection{Reverse Monte Carlo EXAFS} \label{sec:rmc}
Reverse Monte Carlo EXAFS is the process by which atomic coordinates are optimized in an iterative process as a function of the agreement between the theoretical and experimental EXAFS. 
In each iteration, the coordinates are updated, and theoretical EXAFS is calculated. 
The updated coordinates are accepted if the agreement between the theoretical EXAFS and the experimental EXAFS is improved. 
If the updated coordinates provide a worse agreement, then they may be either accepted with some probability or discarded. 
This is a stochastic process, but the probability of acceptance helps to avoid local minima, and after many thousands of iterations (up to several CPU-days), the theoretical EXAFS will converge with the experimental EXAFS given that the initial structure is appropriate.

Here we employed EvAX $5.17$ \cite{timoshenko2014exafs} with FEFF $8.5$ lite \cite{Rehr2010Parameter-freeFEFF9} to implement RMC-EXAFS for the experimental data in \cite{frenkel_view_2001}. 
For each temperature, the \textit{fcc} Pt unit cell was expanded into a ($5\times5\times5$) supercell, and periodic boundary conditions were used. 
The experimental EXAFS data were previously processed in Athena \cite{Ravel:ph5155}. 
The complete EvAX input file, including FEFF simulation parameters, is provided in the SI.

\subsection{EXAFS from ML MD Simulations}
This section is continued from the Methods presented in the main text.
For bulk, each frame is associated with its own MSRD and R. 
Therefore, the EXAFS calculations should be performed for a sufficient number of atoms in each frame so that the averaged MSRD and R between these atoms represent the MSRD and R of the entire frame. 
To find this number of atoms per trajectory, we investigated clusters of atoms from the center of the trajectory at some cluster radius. 
The MSRD and R versus cluster radius are calculated, and the cluster radius at which the MSRD and R converge to the entire frame MSRD and R is $4$ \AA{}, which corresponds to approximately 18 atoms. 
The 18 atoms in the center are designated as absorbers, and the simulation box is truncated to $12$ \AA{} in each direction (truncating from $4000$ atoms to $\sim 444$ atoms to fit inside the FEFF calculation). 
Finally, for each absorbing atom, a FEFF input file is generated using the FEFF simulation parameters in the main text, resulting in $\sim 1782$ FEFF calculations per trajectory. 
The total number of FEFF calculations is then separated into 4 working directories per trajectory, where a cluster node can handle each directory, and each FEFF calculation proceeds sequentially utilizing a total of 36 cores. 
All $15$ temperature steps per MLFF can be calculated in approximately $4$ hours of walltime.

\section{Additional Bulk Results}
\subsection{Validation of Bulk MLFFs with MD-EXAFS}
\begin{figure*}
\centering
\includegraphics[width=\textwidth]{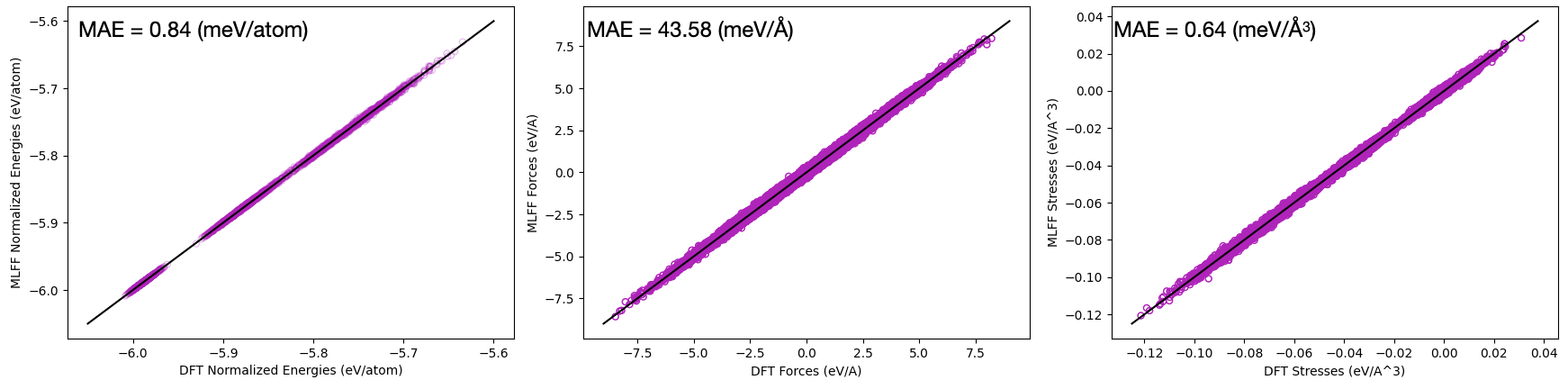}
\caption{Parity results for the PBE bulk Pt MLFF to the energy, force, and stress labels over the entire DFT training set from the TM23 data set in \cite{owen2023complexity}. MAEs are provided in the upper left panel for each label.}\label{fig:pbe}
\end{figure*}

\begin{figure*}
\centering
\includegraphics[width=\textwidth]{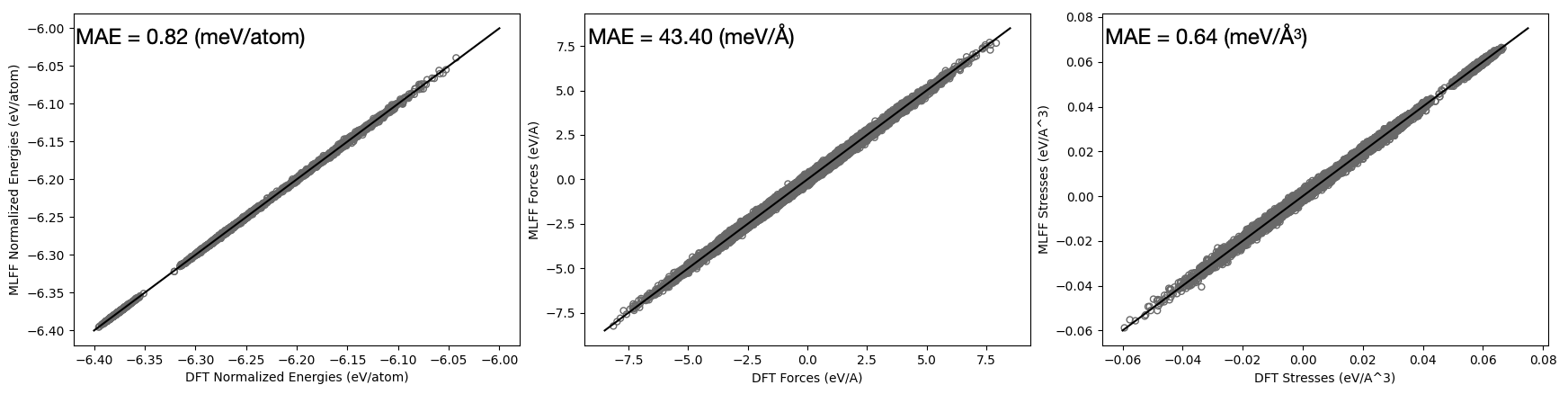}
\caption{Parity results for the PBEsol bulk Pt MLFF to the energy, force, and stress labels over the entire DFT training set from the TM23 data set in \cite{owen2023complexity}. MAEs are provided in the upper left panel for each label.}\label{fig:pbesol}
\end{figure*}

\begin{figure*}
\centering
\includegraphics[width=\textwidth]{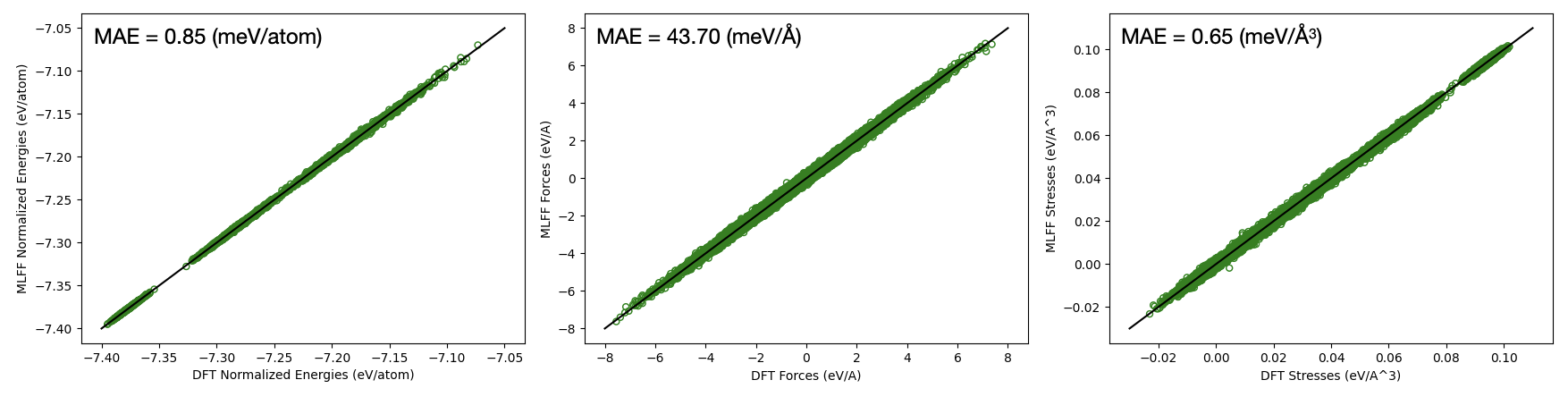}
\caption{Parity results for the LDA bulk Pt MLFF to the energy, force, and stress labels over the entire DFT training set from the TM23 data set in \cite{owen2023complexity}. MAEs are provided in the upper left panel for each label.}\label{fig:lda}
\end{figure*}

We first investigated a set of MLFFs trained exclusively to describe bulk Pt systems. 
Since it is well understood that MLFFs are only as physically descriptive and accurate as the DFT functional with which they are trained, benchmarking begins with respect to the various DFT functionals applied to Pt. 
Here, PBE, PBEsol, and LDA exchange-correlation functionals were employed to calculate training labels from the TM23 dataset \cite{owen2023complexity}, from which three distinct MLFFs were trained. 
Each MLFF was constructed according to the procedure in the main text, where the models observe the same training structures but with different force, energy, and stress labels calculated using each functional. 
The training data for each MLFF here was obtained from \cite{owen2023complexity}, containing $3000$ frames for Pt across three temperatures ($0.25$, $0.75$, and $1.25\cdot T_{melt}$, where $T_{melt}$ is the experimental melting temperature of Pt). 
This set of $3000$ frames were recomputed using each exchange-correlation functional, and an independent MLFF was trained on each collection of energy, force, and stress labels.
Parity results for each of the MLFFs' predictions across their respective training sets from the PBE, PBEsol, and LDA functionals are provided in Fig. \ref{fig:pbe}, Fig. \ref{fig:pbesol}, and Fig. \ref{fig:lda}, respectively.
An important observation is made from these comparisons, in that the choice of exchange-correlation functional does not change the accuracy of the MLFF predictions.

The experimental EXAFS used for comparison was collected from a bulk Pt foil at $200$, $300$, $473$, and $673$ K and analyzed in \cite{SanchezNonBulk}. 
We observed $10$ ns trajectories (each containing $4000$ atoms) at $15$ incremental temperatures for each functional – the equivalent of $39,000$ years of DFT calculations in only $1$ hour of wall time, as each simulation is run in parallel, each using $1$ A100 GPU and obtained an average performance of $250$ ns/day. 

\begin{figure*}
\centering
\includegraphics[width=0.8\textwidth]{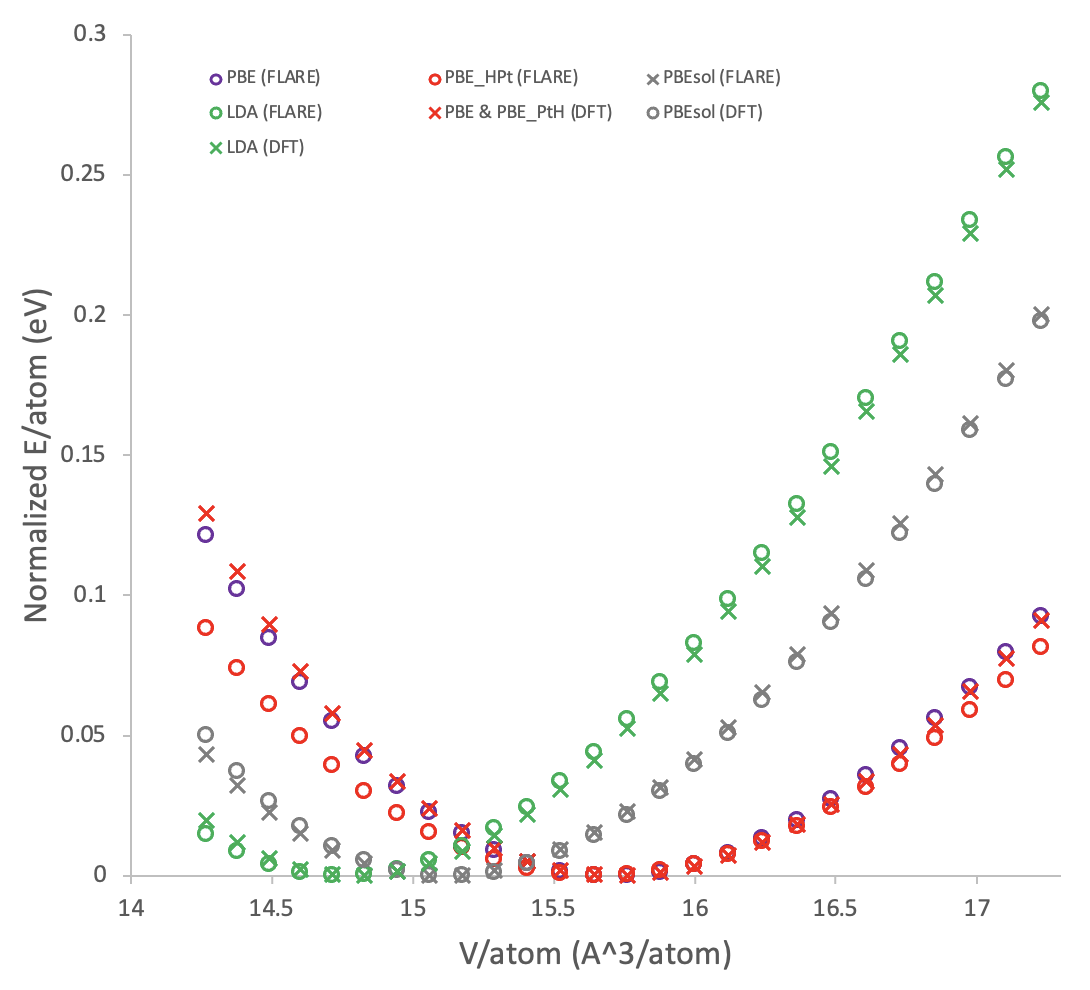}
\caption{Comparison of the functional-specific MLFFs and DFT predictions using the PBE, PBEsol, and LDA exchange-correlation functionals for energy versus volume predictions with FLARE (open circles) and DFT (crosses) predictions, normalized by the minimum energy of each DFT prediction.}\label{fig:mlff_eos}
\end{figure*}

\begin{figure*}
\centering
\includegraphics[width=0.8\textwidth]{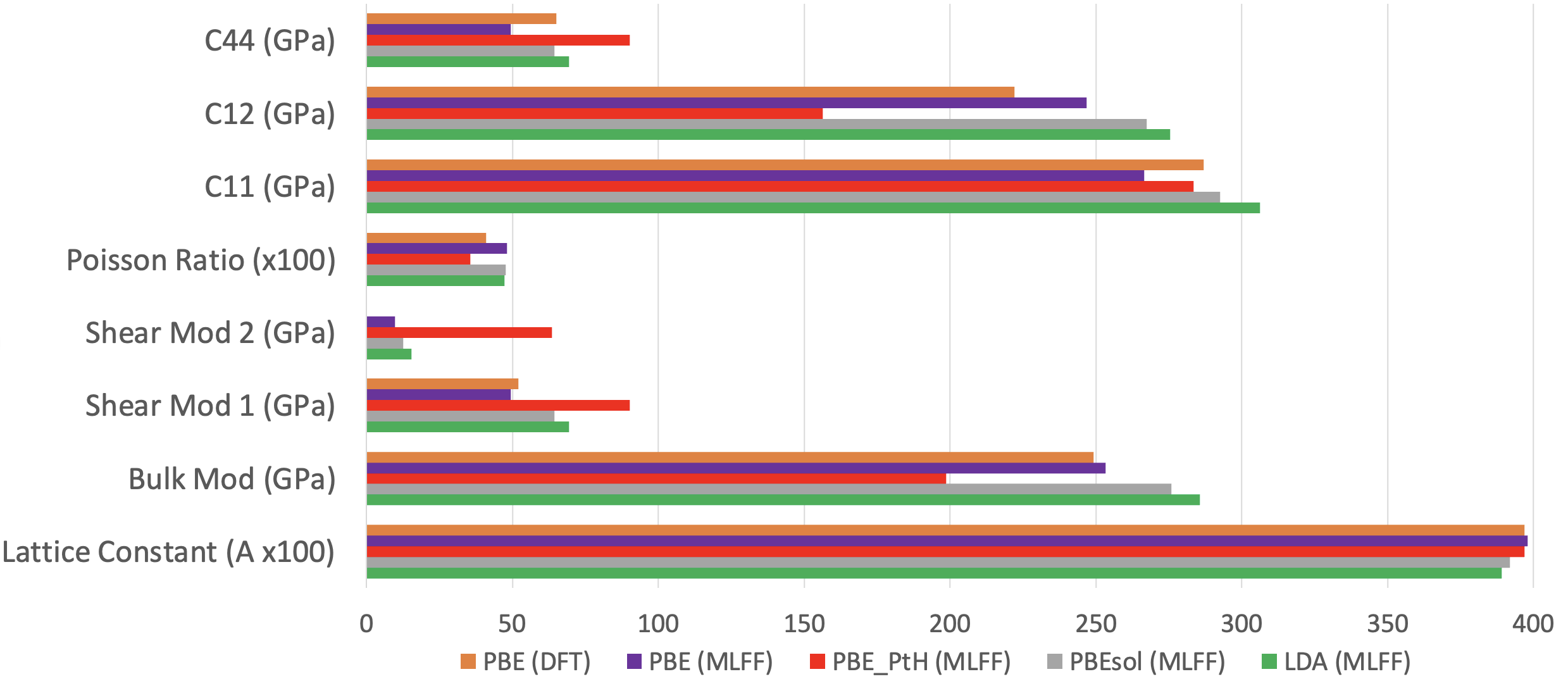}
\caption{Comparison of MLFF and DFT (only at the PBE level) predictions on bulk modulus, elastic constants, and lattice constant.}\label{fig:mlff_elastic}
\end{figure*}

\begin{figure*}
\centering
\includegraphics[width=\textwidth]{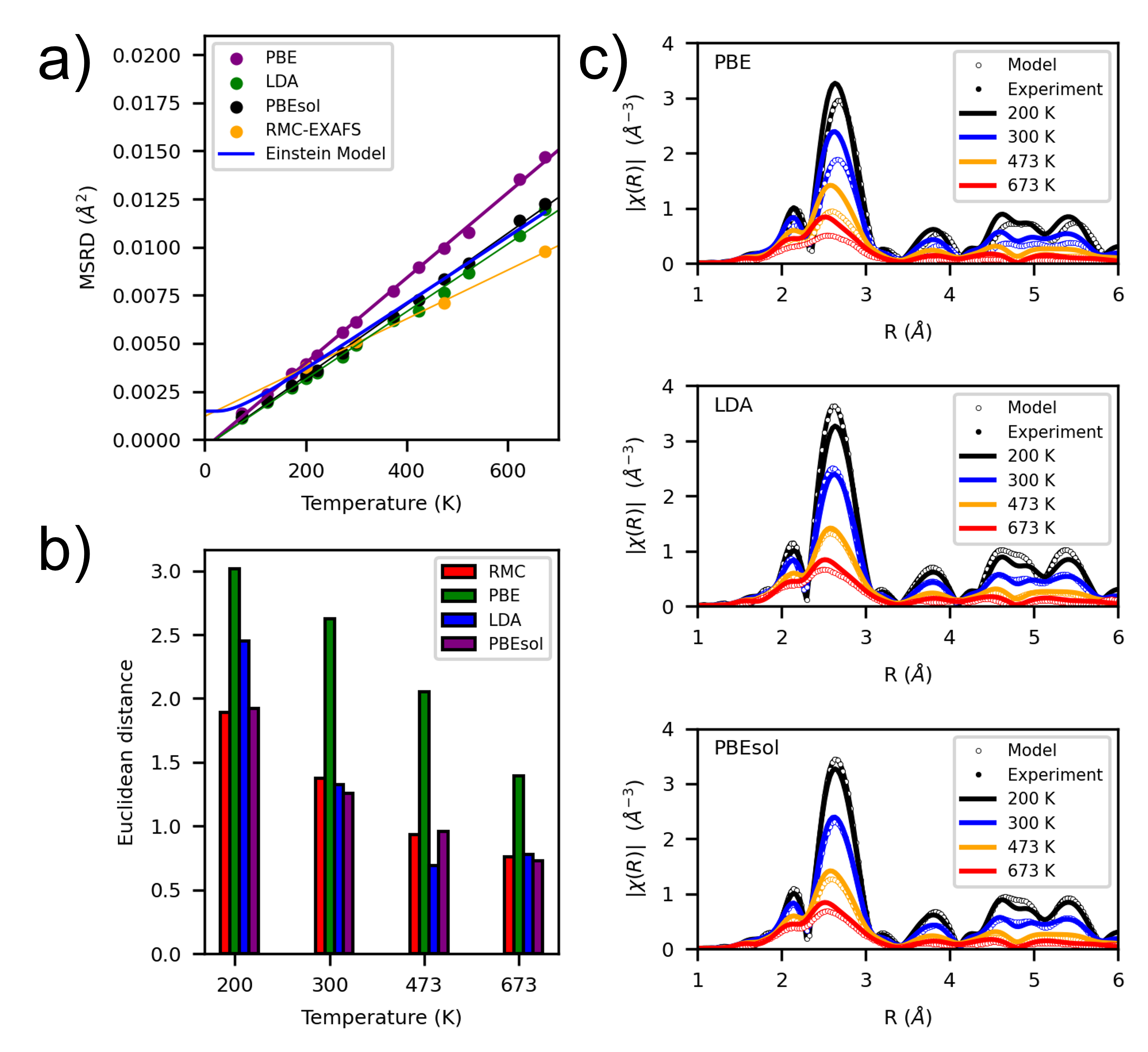}
\caption{MLFFs trained using PBE, PBEsol, and LDA exchange-correlation functionals are benchmarked against experimental FT-EXAFS collected from bulk Pt foil at 200, 300, 473, and 673 K \cite{SanchezNonBulk}. (\textbf{a}) MSRD of the average Pt-Pt interatomic distance from the MLFF MD trajectories as a function of temperature, which is compared to RMC-EXAFS and the Einstein Model. (\textbf{b}) Euclidean distance metrics to illustrate the agreement between MD-EXAFS obtained from each MLFF MD trajectory (\textbf{c}) FT-EXAFS calculated for each MLFF MD trajectory compared to the experimental EXAFS at the same temperatures. 
All FT-EXAFS is taken between 3 and 12 Å$^{-1}$ with a k-weighting of 2 and using the Kaiser-Bessel window function}
\label{fig:bulk_validation}
\end{figure*}

Traditional static benchmarks (\emph{e.g.} lattice and elastic constants) for each MLFF are also provided in Fig. \ref{fig:mlff_eos} and Fig. \ref{fig:mlff_elastic}. 
Based on the over- and under-estimation of the $0$ K lattice constant of Pt ($3.90$ -- LDA, $3.92$ -- PBEsol, and $3.97$ \AA{} -- PBE) compared to $3.92$ \AA{} -- expt., it is expected that PBEsol and LDA will provide the best comparison to experiment due to this baseline agreement. 
However, performance at higher temperatures relies on the agreement between the experimental and theoretical force constants, which can be inferred from the slopes of the MSRD as a function of temperature \cite{PhysRevB.48.585} in Fig. \ref{fig:bulk_validation}a. 
A trajectory analysis was performed (procedure detailed in the Methods section in the main text) to obtain the average R and MSRD of the atomic positions over the MD trajectories. 
These values are physically meaningful and are commonly extracted from EXAFS data via conventional and advanced fitting methods \cite{kas2021advanced}. 
The MSRD is due to the static (heterogenous bonding environments) and dynamic (atomic displacement and thermal vibration) bonding disorder.
 
To benchmark R and MSRD from the MLFF trajectories, the experimental values needed to be known; thus, the Reverse Monte Carlo EXAFS (RMC-EXAFS) method was also employed. 
RMC-EXAFS is the process by which 3D coordinates are optimized so that the resulting theoretical spectrum is as close as possible to the experimental spectrum. 
The resulting 3D coordinates represent a configurational average by which R and MSRD can be extracted. 
While RMC-EXAFS is powerful, it can only be applied reliably to a material where the coordination numbers (CNs) are known, for example, bulk Pt, which assumes the fcc crystal structure (CNs = $12$).

The results for R and MSRD versus temperature are shown in Fig. \ref{fig:bulk_validation}(a). 
As expected, the PBE and LDA MLFFs overestimate and underestimate the interatomic distance, respectively. 
This was expected based on the static benchmark of the lattice constant at $0$ K and is due to the underbinding of PBE and the overbinding of the LDA functionals, respectively. 
PBEsol more closely matches the R obtained by RMC-EXAFS due to the corrections implemented in the functional, providing exact agreement with the $0$ K lattice parameter. 
The slope of MSRD versus temperature is inversely proportional to the force constant, while the y-intercept is correlated with the amount of static disorder present in the system. 
Notably, the y-intercept of the experimental data is not exactly zero, which indicates that some static disorder is present, possibly due to the manufacturing process (\emph{e.g.}, cold-rolling of the Pt-film, which creates grain boundaries and defect sites that are not annealed afterward).
The slope of the PBE MLFF is larger than that of the LDA MLFF, indicating a `softer' force constant, with PBEsol also matching this behavior. 
Regardless of their agreement with the equilibrium lattice constant of Pt, all of the MLFFs employed here overestimate the MSRD with increasing temperature. 
The Einstein model of the disorder, which does well for modeling thermal disorder in bulk metals \cite{PhysRevB.20.4908}, is also plotted for comparison purposes. 
This model lacks the intrinsic static disorder of the experimental foils but accounts for quantum-mechanical disorder that arises from zero-point vibrational energy at low temperatures. 

EXAFS was calculated for a subset of MD frames which were then averaged to provide the MD-EXAFS for each functional across temperatures. 
For comparison with experimental EXAFS, the Euclidean distance was taken between the experimental spectra and the FT-MD-EXAFS at each temperature and the same was done for RMC-EXAFS. 
These comparisons are presented in Fig. \ref{fig:bulk_validation}(b), where the height of the bar is proportional to a decrease in agreement. 
RMC-EXAFS was expected to be in the best agreement with the experiment by the nature of the procedure; however, PBEsol and LDA agreed best with experiment, followed by RMC-EXAFS, and then PBE. 
The FT-EXAFS peaks in Fig. \ref{fig:bulk_validation}(c) can be roughly assigned to the first (between $2-3$ \AA{}), second (between $3.5$ and $4$ \AA{}), and third and fourth nearest neighbor scattering paths (between $4.5$ and $6$ \AA{}). 
Agreement beyond the first peak is notable because these peaks arise from many-body interactions. 
The excellent agreement observed between the MD-EXAFS calculated from the PBEsol trajectory, in combination with the agreement between the MSRDs calculated on the PBEsol and LDA trajectory with the Einstein model allows us to conclude that the difference between the reduced slope of the RMC-EXAFS MSRDs is due to intrinsic disorder that is not captured by the MD simulation.

In summary, MLFFs for describing bulk Pt were investigated as a function of the underlying exchange-correlation functional (PBE, PBEsol, and LDA). 
Trajectory analysis (R and MSRD) and FT-MD-EXAFS were used to benchmark the MD simulations across temperatures ranging from $200$ to $673$ K. 
In all cases, R scales with the prediction of the lattice constant at $0$ K by each functional. 
Dynamic disorder due to finite temperature is over- or under-estimated in proportion to the force constant, or ‘stiffness,’ of the functional, but all of the MLFFs constructed here overestimate thermal disorder relative to experiment. 
Peaks due to multiple scattering effects in the FT-MD-EXAFS, however, are reproduced by all three MLFFs, where the peak amplitude scales as a function of functional stiffness and peak location shifts according to the lattice constant at $0$ K.

\subsection{Benchmarking the HPt MLFF}
\begin{figure*}
\centering
\includegraphics[width=\textwidth]{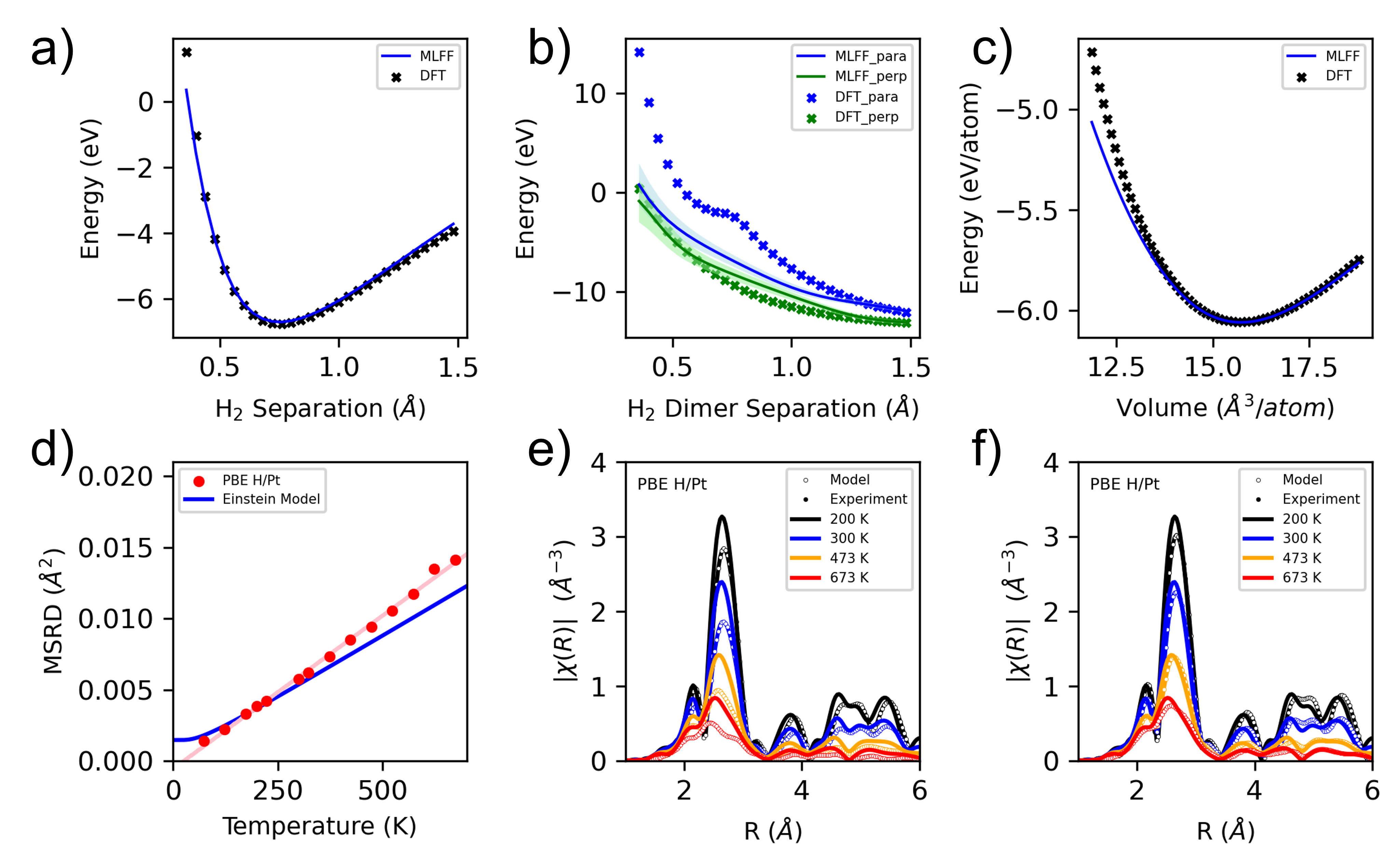}
\caption{Benchmark results for the HPt MLFF with static DFT calculations and MD-EXAFS. (\textbf{a}) Total energy vs. separation of an H$_2$ molecule. The equilibrium bond length of 0.74 Å predicted by DFT is replicated by the MLFF. (\textbf{b}) Total energy vs. H$_2$ dimer separation in parallel and orthogonal configurations. (\textbf{c}) Total energy vs. volume for a periodic bulk cell of Pt. (\textbf{d}) Average MSRD of the Pt-Pt interatomic distance as compared to that predicted by the Einstein model. (\textbf{e}) FT-EXAFS calculated for each MLFF MD trajectory as compared to the experimental EXAFS at the same temperature. (\textbf{f}) MD-EXAFS compared to the experiment after implementing the temperature matching scheme described in the main text.}
\label{pth_validation}
\end{figure*}

\begin{figure*}
\centering
\includegraphics[width=\textwidth]{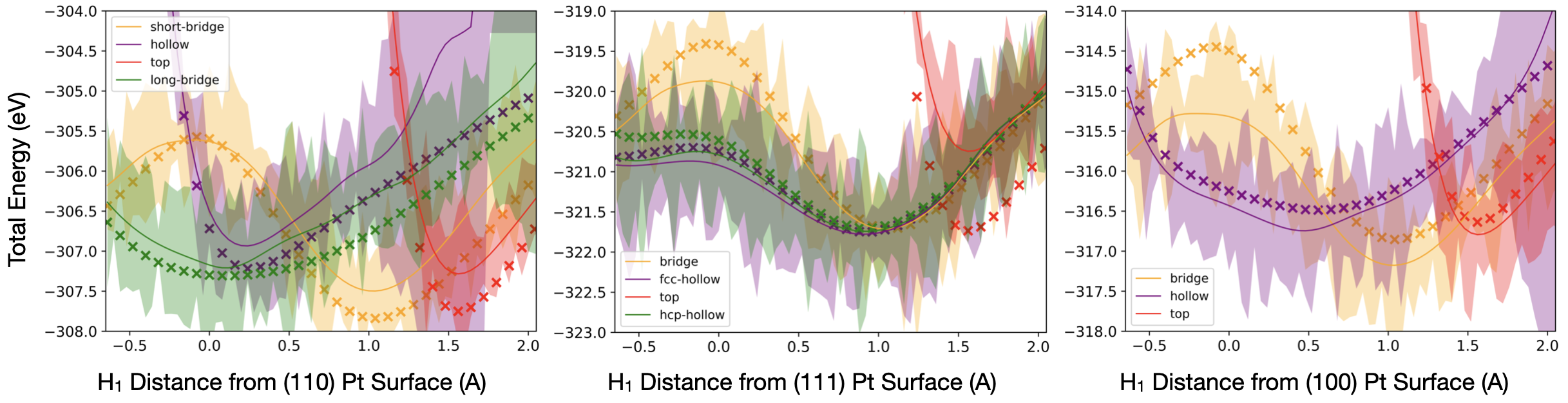}
\caption{H adsorption barriers in terms of total energy on (\textbf{left}) Pt(110), (\textbf{middle}) Pt(111), and (\textbf{right}) Pt(100) as a function of site and vertical height in the normal direction of the surface Pt atoms. All calculations are single point evaluations. The 99\% confidence interval of each prediction is shown for each prediction. No relaxation is performed. Crosses represent DFT predictions, while the line corresponds to those from the MLFF.}
\label{fig:h1_ads}
\end{figure*}

\begin{figure*}
\centering
\includegraphics[width=\textwidth]{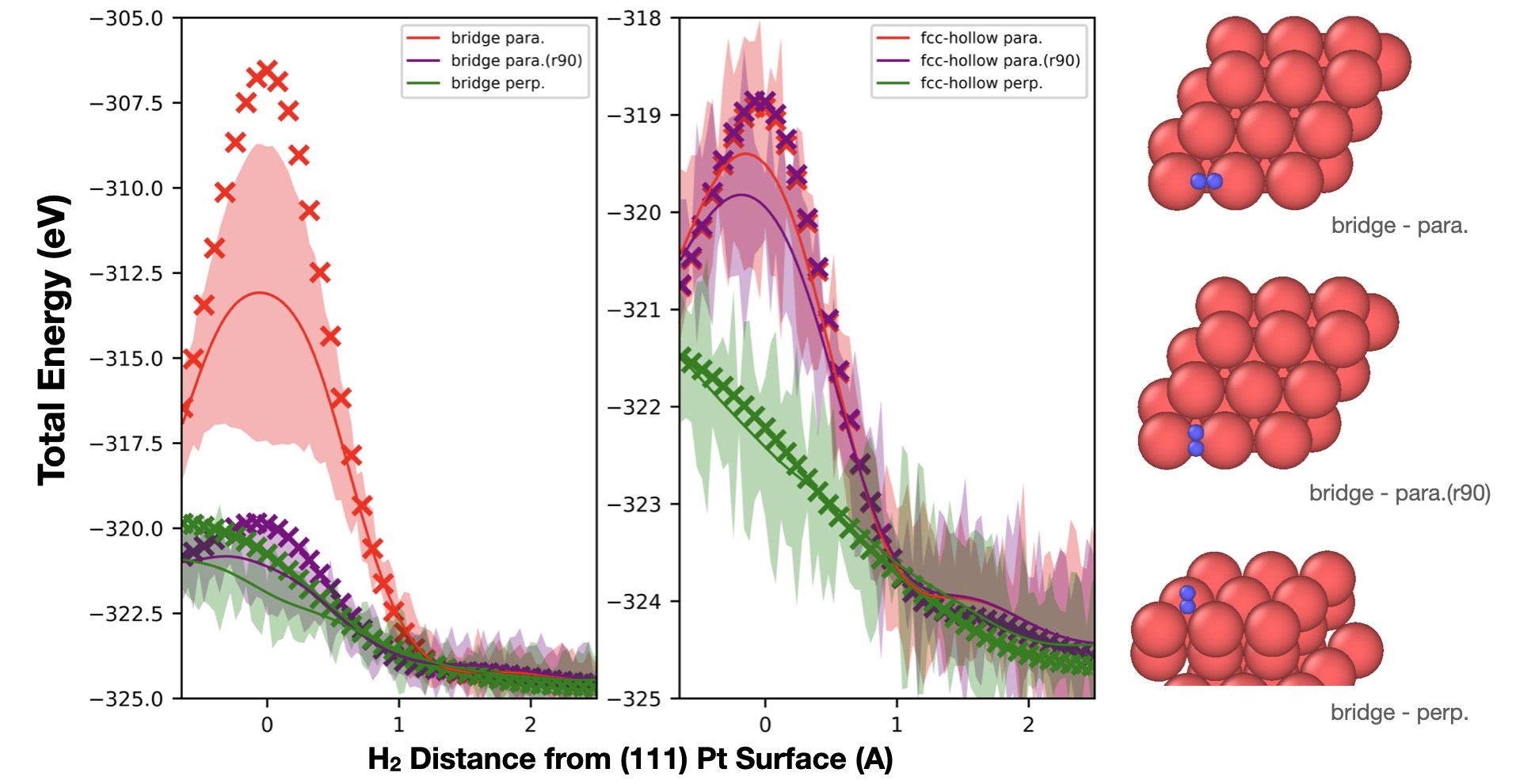}
\caption{H$_2$ adsorption barriers in terms of total energy on Pt(111) as a function of adsorption site, (\textbf{left}) bridge and (\textbf{right}) \textit{fcc}. All calculations are single point evaluations. The 99\% confidence interval of each prediction is shown for each prediction. A schematic of the bridge interaction and orientation of the H$_2$ molecule. Crosses represent DFT predictions, while the line corresponds to those from the MLFF.}
\label{fig:h2_ads}
\end{figure*}

The overarching goal of this work was to create an MLFF that can simultaneously describe reactions over Pt nanoparticle systems and the response of the particle interfaces and structure to such stimuli. 
Thus, an MLFF was trained to describe Pt-bulk, Pt-H, and H-H interactions. 
Briefly, these interactions were captured via augmentation of the bulk PBE training set from the TM23 work with Pt NPs, slabs, H$_2$ gas, and the NP and slab systems exposed to H$_2$, as described in the Methods section in the main text. 
Despite the PBE exchange-correlation functional not displaying the best agreement between MD-EXAFS and experimental EXAFS in the previous section, this exchange-correlation functional was chosen for the reactive system due to over-prediction of barrier heights by PBEsol for dissociative adsorption and recombinative desorption of H$_2$ on the Pt(111) facet \cite{WELLENDORFF201536}. 
Moreover, previous work on the H/Pt(111) system demonstrated excellent agreement with respect to the activation energy of these reactions via Arrhenius analysis \cite{Vandermause2022}, giving further creedence to our choice of the PBE functional to provide accurate force, energy, and stress labels for this reactive system. 
The HPt MLFF is created according to the procedure outlined in the Methods section in the main text and in \cite{Vandermause2022} and \cite{Johansson2022Micron-scaleLearning}. 
The bulk data from the TM23 dataset in \cite{owen2023complexity} was also employed.
Following the collection of training data and building the final MLFF, various static DFT benchmarks were performed to assess the robustness of the potential across Pt-Pt, Pt-H, and H-H interactions. 
These results are presented in Fig. \ref{pth_validation}, where excellent agreement is observed for the separation of H$_2$, producing the equilibrium bond length of $0.741$ \AA{} \cite{doi:10.1021/ja077011d}, with the MLFF accurately predicting energy as a function of separation as compared to DFT. 
The dimer interactions between H$_2$ molecules is also considered, as was done in the original HPt FLARE work \cite{Vandermause2022}, where the qualitative agreement of an energetic barrier at low separation distances is observed. 
Quantitative agreement of the barrier height is not present, but the presence of a barrier is important as the system will not collapse on itself, \emph{e.g.} predict non-physical structures with overlapping atoms to be low-energy, while in production MD. 
This is an important observation with respect to the stability of the potential, as was mentioned in our previous work concerning H/Pt($111$) at the scale of $0.5$ trillion atoms \cite{Johansson2022Micron-scaleLearning}. 
The bulk Pt system is also considered with respect to bulk modulus and elastic constants. 
Excellent agreement is observed via the prediction of energy per atom as a function of volume, with a slight deviation appearing at small volumes. 
This is plausibly due to the inclusion of nanoparticles in the training data. 
This is a plausibly reasonable explanation as atomic environments at the center of the nanoparticle may reflect bulk \textit{fcc} symmetry, but the local environment is distorted relative to proper bulk packing. 
The barriers for H and H$_2$ adsorption were also considered, as shown in Fig. \ref{fig:h1_ads} and Fig. \ref{fig:h2_ads}, respectively.
From these static comparisons, it can be concluded that the MLFF describes H$_2$-H$_2$ and H$_2$-Pt interactions with near DFT accuracy, providing qualitatively correct barriers, limiting the MLFF to enter non-physical domains during production MD.

The results of trajectory analysis are presented in Fig. \ref{pth_validation}(d) as compared to the Einstein model for the bulk Pt system to establish consistency with the PBE, PBEsol, and LDA MLFFs explored in the previous section. 
MD-EXAFS was calculated from a subset of MD trajectory frames and averaged at the experimental temperatures. 
The performance, based on the Euclidean distance from the experimental data, is approximately the same as the bulk MLFF using the PBE functional. 
The EXAFS comparison is provided in Fig. \ref{pth_validation}(e). 
Briefly, the peaks of the FT-MD-EXAFS are underestimated due to the overestimation of MSRD with temperature, correlating with a weaker force constant compared to the real Pt sample, and the average position of the first peak is shifted higher in R.

\subsubsection{Temperature-Matching Protocol}
Before initiating the nanoparticle simulations described in the main text, we first established a protocol to correct for the overestimation of MSRD via a `temperature-matching' method described in the Methods section in the main text. 
The principle of temperature matching is that the MSRD versus temperature plot only differs from the Einstein model by a constant multiple of the slope due to a difference in force constant. 
Therefore, we can determine an empirical relationship between the MD temperature to the experimental temperature provided by the difference in MSRD. 
We align the MD temperature with the Einstein model, which provides MSRDs that best fit the experimental data. 
From this alignment, we determine that the HPt PBE-MLFF must survey dynamics at $196.27$, $269.96$, $397.45$, and $544.84$ K to better describe the MSRD of experimental bulk Pt at $200$, $273$, $473$, and $673$ K, respectively for the bulk simulations, and $255.88$ K for comparison to $300$ K experimental data for the NPs. 
The result of MD-EXAFS after this temperature matching scheme is presented in Fig. \ref{pth_validation}(f). 
The agreement in peak height is markedly better at high temperatures across both pairwise and many-body scattering paths. 
Thus we can conclude that this method is sufficient for correcting for the overestimation of MSRD due to the intrinsic underbinding of the PBE functional, which will be important in simulating the subsequent NPs at relevant experimental conditions.

Based on the results presented in this section, we have established confidence that the HPt PBE-MLFF can describe Pt-Pt, Pt-H, Pt-H$_2$, and H$_2$-H$_2$ interactions with quantum-mechanical accuracy and that the MSRD-based temperature matching scheme allows for more appropriate simulation of these systems at experimentally relevant conditions.

\subsection{MLFF Training via Active Learning}
As is discussed in more detail in the Methods section of the main text, only the bulk-Pt, and Pt($111$), H/Pt($111$), Pt$_{55}$, and Pt$_{147}$ systems with and without exposure to H$_2$ were considered in the active learning simulations performed here.
These systems and the active learning simulations are summarized in Table \ref{tab:active}.
The simulation time ($\tau_{\text{sim}}$), temperature (Temp. (K)), number of DFT calls ($N_{\text{DFT}}$), number of atoms ($N_{\text{atoms}}$) and total wall-time ($t_{\text{wall}}$) are recorded, along with the hardware ($N_{cpus}$) employed to carry out the simulations. 
Summarily, only $14.6$ days of walltime are required to survey a total of $\sim 2.1$ ns of dynamics, where $893$ DFT calls were made. 

By running these simulations in parallel, only $6$ days of computational wall-time were required, using a total of $29$, $32$-core CPUs on the Harvard Cannon cluster.
Considering the total number of atoms, systems, and time-steps considered, our active-learning workflow results in an incredible acceleration (on the order of centuries) relative to a pure \textit{ab initio} study of the same systems.

\subsection{NN Surrogate Model}
\begin{figure*}
\centering
\includegraphics[width=\textwidth]{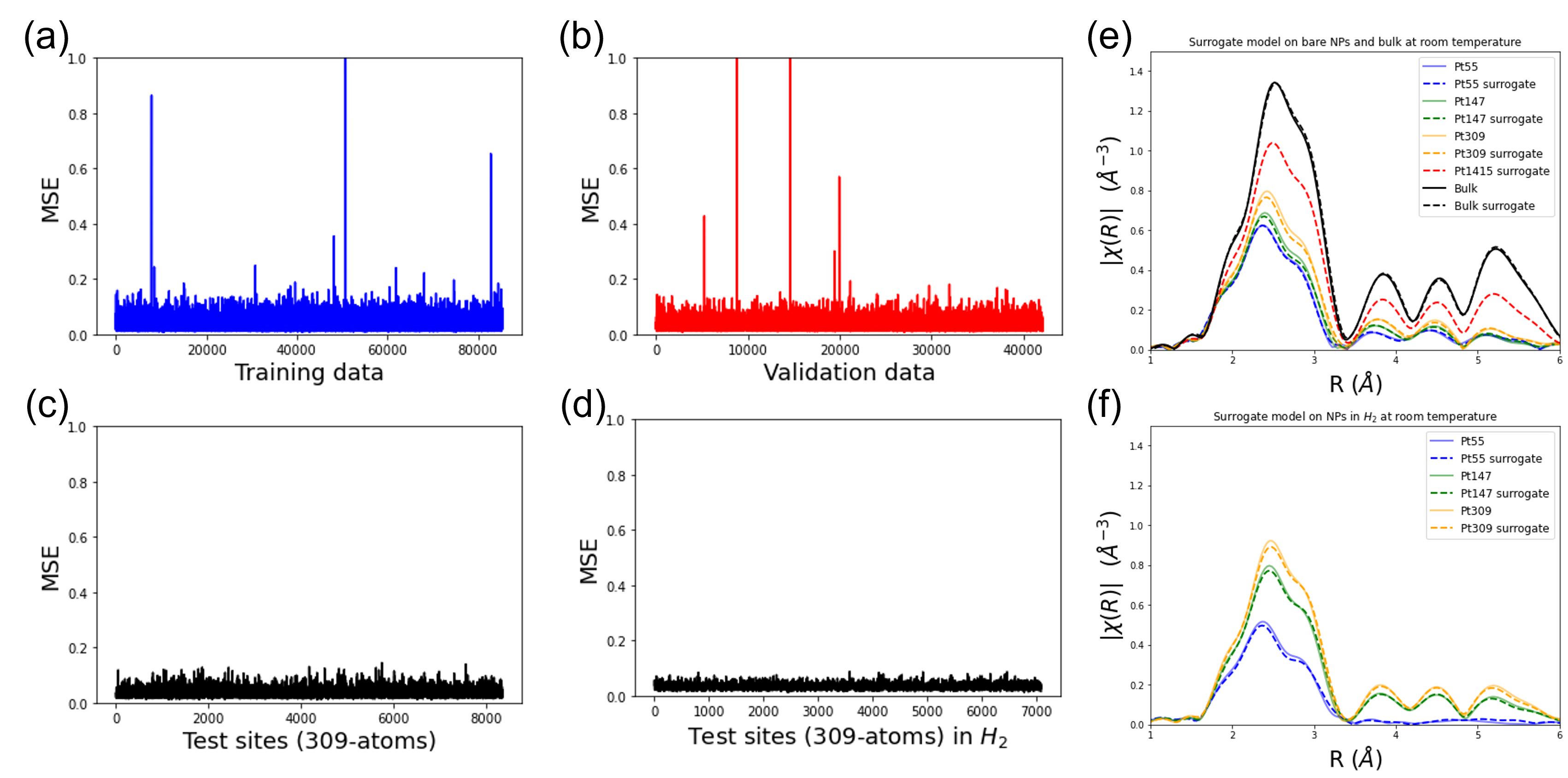}
\caption{Validation and results of NN-surrogate training and testing. The Mean Squared Error for the predictions made on the training (\textbf{a}), validation (\textbf{b}), and test (\textbf{c,d}) sets. The FT-EXAFS of the true and surrogate predictions of bare (\textbf{e}) and hydrogen-exposed (\textbf{f}) NPs.}
\label{fig:NN_surg_val}
\end{figure*}

A convolutional neural network (CNN) was trained using Keras to map site-specific discretized total radial distribution functions to site-specific EXAFS calculations. 
The CNN consists of an input layer of size 240 connected to a reshape(3, 80) layer, followed by two convolutional layers with kernel = 2 and output size = 32, a flatten layer, two dense layers of 600 neurons each, and finally an output layer of 201 neurons. 
The intermediate layers are activated with Tanh. 
The training was performed using the Adam optimizer with learning rate = 0.0001 and a batch size of 64 over 483 epochs. 
The training data set contains a total of 127,475 sites, which come from the cuboctahedral 55 and 147 atom NPs in both bare and H$_{2}$ conditions from their respective simulations at 256 K, as well as the bulk sites from all bulk simulations using the HPt PBE MLFF. 
During training, 33\% of the sites were reserved for the validation set. 
The best NN-surrogate is presented in Fig. \ref{fig:NN_surg_val}.

\section*{References}
\bibliography{bib.bib}

\section{Supplementary Tables and Figures}
\begin{table*}[t]
\centering
\begin{tabular}{|c|c|c|c|c|c|c|}
\multicolumn{1}{c}{\bf System}    & \multicolumn{1}{c}{\bf Temp. (K)}   &  \multicolumn{1}{c}{\bf $\tau_{\textrm{sim}}$ (ps)} & \multicolumn{1}{c}{\bf   $\tau_{\textrm{wall}}$ (hr)} & \multicolumn{1}{c}{\bf $N_{\textrm{cpus}}$} & \multicolumn{1}{c}{\bf $N_{\textrm{atoms}}$} & \multicolumn{1}{c}{\bf $N_{\textrm{DFT}}$}\\
\hline
H$_2$ & 1500 & 5.0 & 1.2 & 32 & 54 & 24 \\ 
H$^{\ast}_2$ & 2100 & 10.0 & 3.5 & 64 & 108 & 2 \\ 
Pt(111) & 300 & 4.0 & 1.2 & 64 & 54 & 4 \\ 
Pt-bulk & 1500 & 10.0 & 1.7 & 64 & 108 & 6 \\
H/Pt(111) & 1500 & 3.7 & 61.4 & 64 & 73 & 216 \\
H/Pt(111)$^{\ast}$ & 2100 & 26.5 & 143.1 & 96 & 92 & 323 \\
Pt$_{55}$ NP & 2100 & 1000 & 12.0 & 32 & 55 & 19 \\
Pt$_{147}$ NP & 2100 & 1000 & 11.0 & 144 & 147 & 14 \\
H/Pt$_{55}$ NP & 2100 & 36.8 & 69.0 & 144 & 147 & 181 \\
H/Pt$_{147}$ NP & 2100 & 3.6 & 45.6 & 320 & 309 & 104 \\
\hline
\end{tabular}
\caption{Summary of the FLARE active-learning trajectories for each of the Pt and HPt systems.\label{tab:active}}
\end{table*}

\begin{figure*}
\centering
\includegraphics[width=\textwidth]{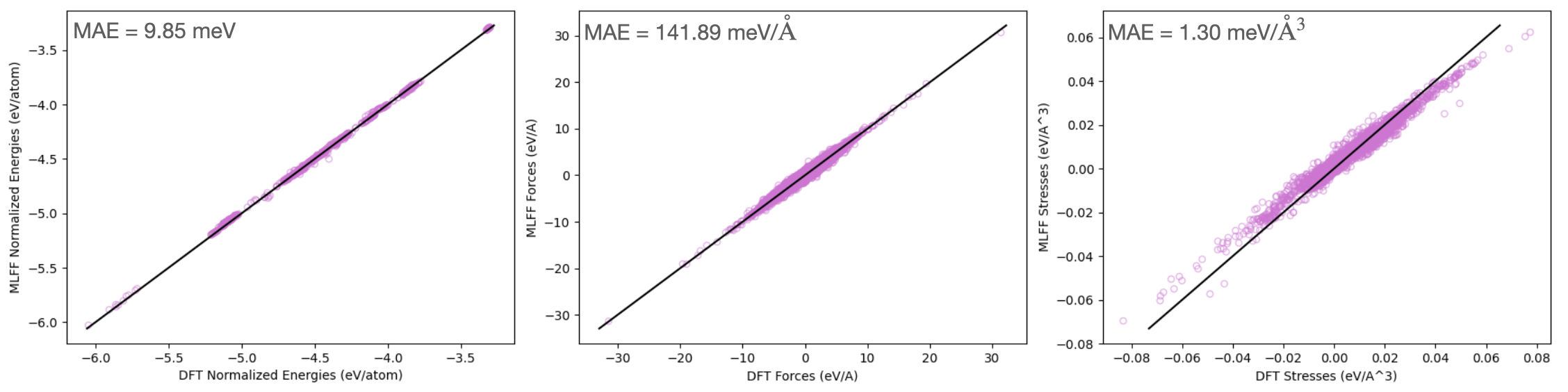}
\caption{Parity results for the HPt MLFF to the energy, force, and stress labels over the entire DFT training set. MAEs are provided in the upper left panel for each label.}
\end{figure*}

\begin{figure*}
\centering
\includegraphics[width=\textwidth]{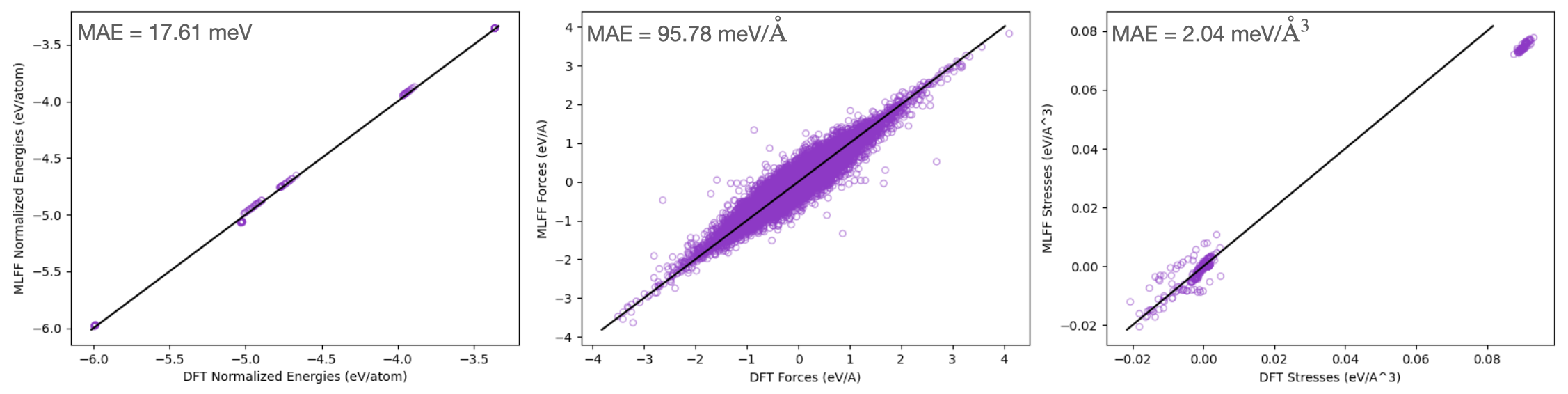}
\caption{Parity results for the HPt MLFF to the energy, force, and stress labels over a test set collected via simulation of DFT-sized cells using the trained MLFF in LAMMPS. MAEs are provided in the upper left panel for each label.}
\end{figure*}

\begin{figure*}
\centering
\includegraphics[width=\textwidth]{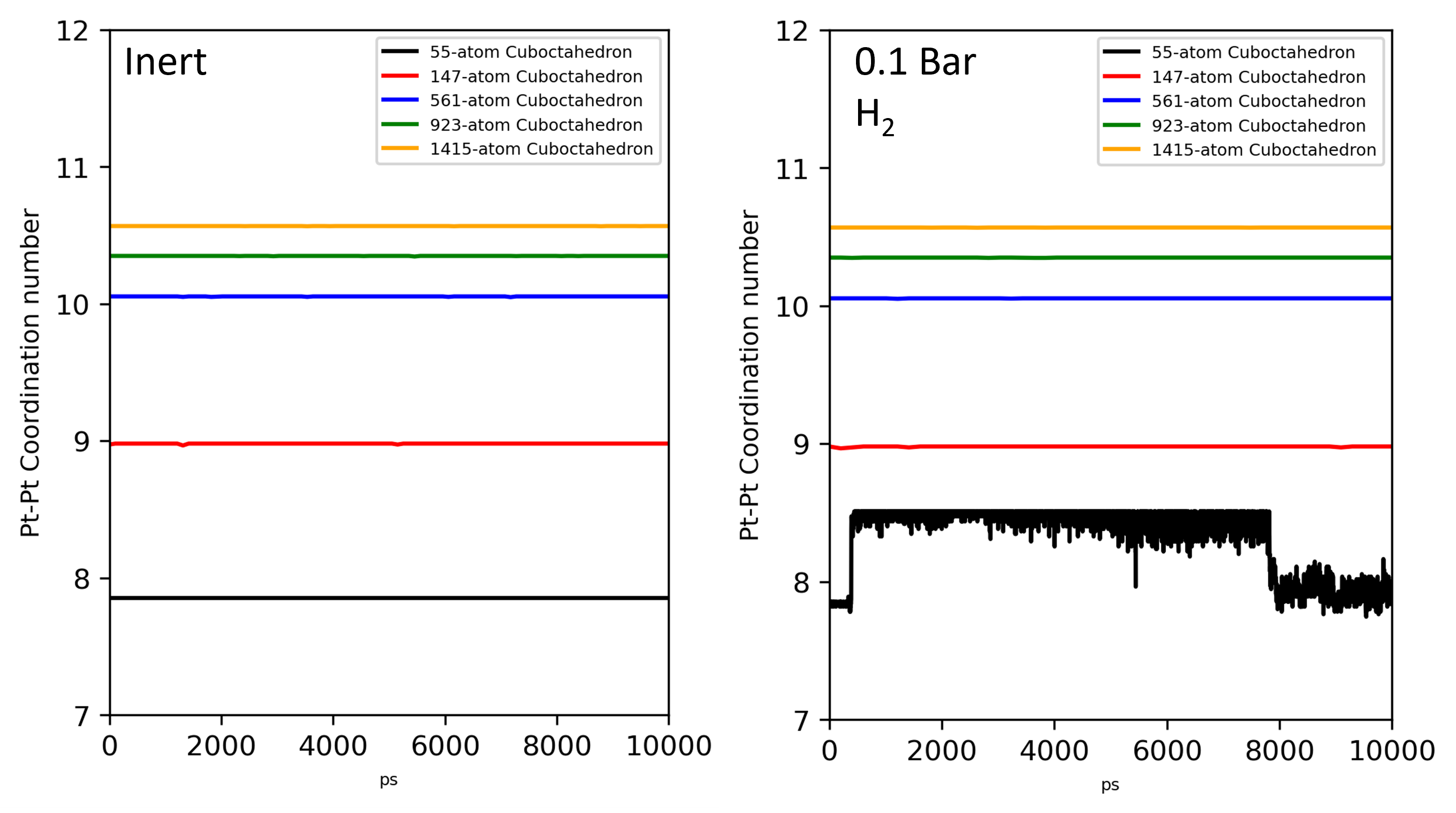}
\caption{Pt-Pt coordination numbers over the cuboctahedral simulation trajectory for the inert (\textbf{left}) and hydrogen-exposed (\textbf{right}) NPs between 55 and 1415 atoms.}
\label{all_cns}
\end{figure*}

\begin{figure*}
\centering
\includegraphics[width=0.6\textwidth]{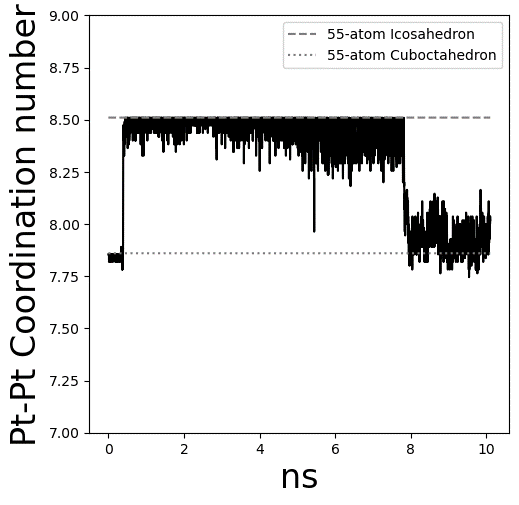}
\caption{Pt-Pt CN of the 55 atom hydrogen-exposed cuboctahedral particle over the simulation trajectory.}
\label{cubo_55_h_cn}
\end{figure*}

\begin{figure*}
\centering
\includegraphics[width=\textwidth]{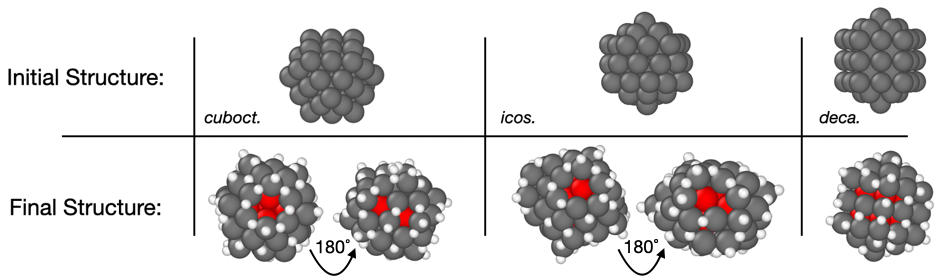}
\caption{Snapshots of the final NP structures after 0.1 bar H$_2$ exposure at 256 K, where multiple rosettes can be observed on the particle surface upon rotation.}
\label{cubo_55_h_cn}
\end{figure*}

\begin{figure*}
\centering
\includegraphics[width=\textwidth]{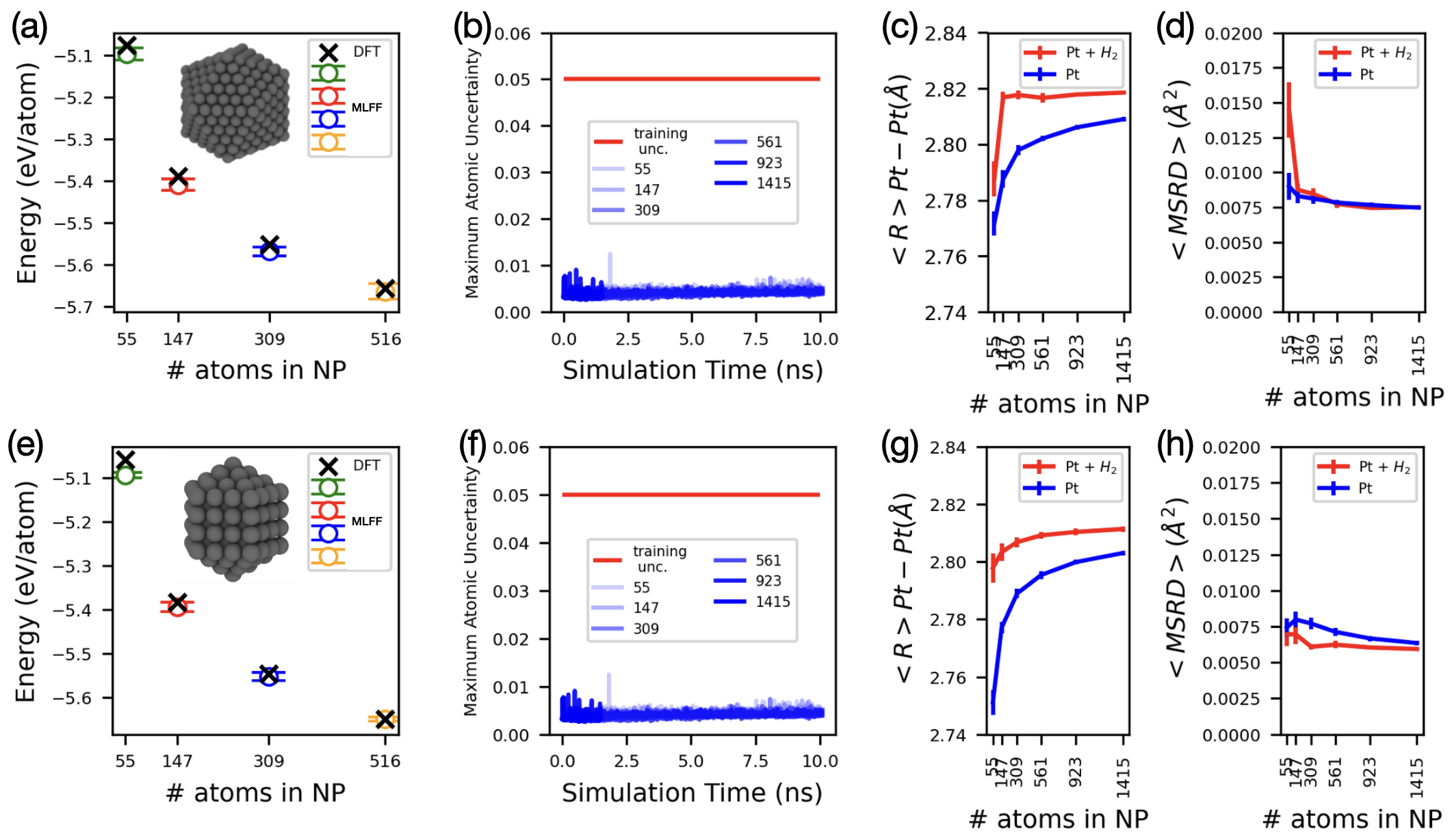}
\caption{(\textbf{Top}) icosahedral and (\textbf{Bottom}) ino-decahedral. (\textbf{a,e}) Comparison of MLFF predicted cohesive energies per atom of Pt icosahedral and ino-decahedral NPs as a function of size. 
Open circles denote MLFF predictions, whereas black crosses are DFT. 
The $99\%$ confidence interval of each prediction by the MLFF is provided. 
(\textbf{b,f}) Maximum atomic uncertainty of ML MD predictions made by the MLFF as a function of MD simulation time. 
(\textbf{c,g}) Average interatomic distance R and (\textbf{d,h}) MSRD for nanoparticles from $55$ to $1415$ atoms with (red) and without (blue) H$_2$ atmosphere. }
\label{fig:extrapolation}
\end{figure*}

\begin{figure*}
\centering
\includegraphics[width=\textwidth]{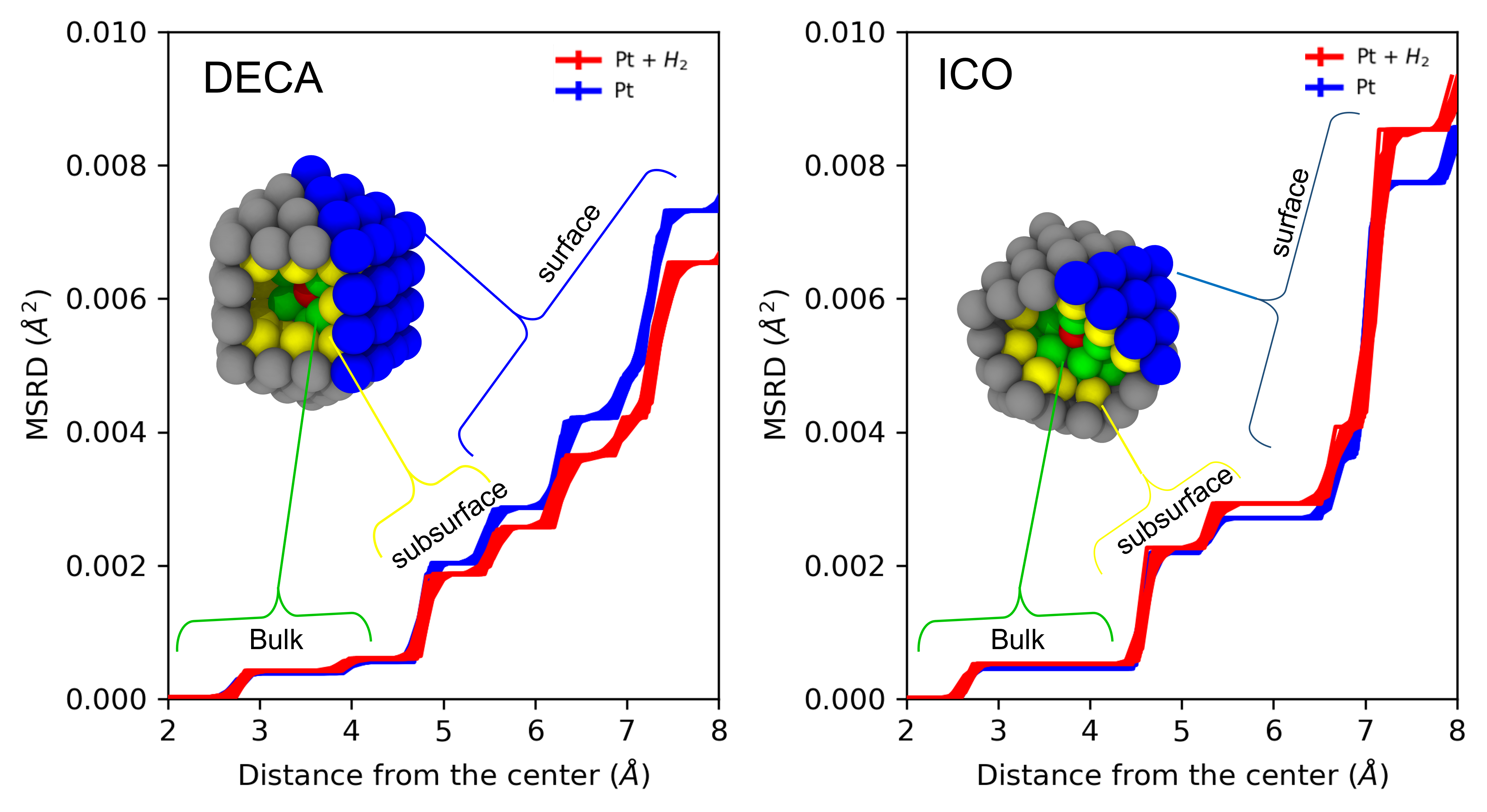}
\caption{The cumulative sum of MSRD contributions from spherical shells of atoms starting from the center of the decahedral (\textbf{left}) and icosahedral (\textbf{right}) for the bare (blue) and H$_2$-exposed (red) simulations. 
The inset shows the particle geometry, and the colors correspond to atoms of similar local environments: `bulk' (green), subsurface (yellow), and surface (blue). 
In both particles, the bulk atom MSRD is about the same. 
Upon H$_2$ exposure, the subsurface and surface atoms in the decahedral particle contribute less MSRD to the cumulative sum, while they contribute more in the icosahedral particle.}
\label{exp_NPs}
\end{figure*}